
\message{NOTE: THE TABLES MUST BE TEX-ED AND PRINTED SEPARATELY.}

\input harvmac.tex

\Title{\vbox{\baselineskip12pt\hbox{CLNS-93/1253}\hbox{IASSNS-HEP-94/2}%
\hbox{YCTP-P31-92}}}
{\vbox{\centerline{ Mirror Manifolds in Higher Dimension}
\centerline{}}}%

\centerline{Brian R. Greene,%
\footnote{${}^\dagger$}{F.R. Newman Laboratory of Nuclear Studies,
Cornell University, Ithaca, NY \ 14853}
David R. Morrison,%
\footnote{${}^\sharp$}{
School of Mathematics, Institute for Advanced Study,
Princeton, NJ \ 08540.
On leave from:  Department of Mathematics, Duke University,
Box 90320, Durham, NC \ 27708-0320}
and M. Ronen Plesser%
\footnote{${}^*$}{Department of Physics,
Yale University, New Haven, CT \ 06511,
and School of Natural Sciences, Institute for Advanced Study,
Princeton, NJ \ 08540}}

\noblackbox

\font\bigrm=cmr10 scaled \magstephalf
\def\inbar{\,\vrule height1.5ex width.4pt depth0pt}
\font\cmss=cmss10 \font\cmsss=cmss8 at 8pt
\def\BZ{\relax\ifmmode\mathchoice
{\hbox{\cmss Z\kern-.4em Z}}{\hbox{\cmss Z\kern-.4em Z}}
{\lower.9pt\hbox{\cmsss Z\kern-.36em Z}}
{\lower1.2pt\hbox{\cmsss Z\kern-.36em Z}}\else{\cmss Z\kern-.4em Z}\fi}
\def\IC{\relax\hbox{$\inbar\kern-.3em{\rm C}$}}
\def\IP{\relax{\rm I\kern-.18em P}}
\def\IQ{\relax\hbox{$\inbar\kern-.3em{\rm Q}$}}
\def\IR{\relax{\rm I\kern-.18em R}}

\def\Tr#1{\hbox{{\bigrm Tr}\kern-1.05em \lower2.1ex \hbox{$\scriptstyle#1$}}\,}

\def\WCP#1#2{\hbox{$\hbox{W\IC \IP}^{#1}_{#2}$}}

\def\e{{\rm e}}

\def\CP#1{\hbox{$\hbox{\IC \IP}^{#1}$}}
\def \p{\partial}
\def\tilde{\widetilde}
\def\vbar{ \,|\, }
\def\ID{\relax{\rm I\kern-.18em D}}
\def\Sym{\mathop{Sym}\nolimits}
\def\lt{ < }
\def\gt{ > }
\def\boldone{\relax{\rm 1\kern-.35em 1}}

\vskip.3in

We describe mirror manifolds in dimensions different from the familiar case
of complex threefolds.  We emphasize the simplifying features of dimension
three and supply more robust methods that do not rely on such special
characteristics and hence naturally generalize to other dimensions.  The
moduli spaces for Calabi--Yau $d$-folds are somewhat different from the
``special K\"ahler manifolds'' which had occurred for $d=3$, and we indicate
the new geometrical structures which arise.  We formulate and apply
procedures which allow for the construction of mirror maps and the
calculation of order-by-order instanton corrections to Yukawa couplings.
Mathematically, these corrections are expected to correspond to calculating
Chern classes of various parameter spaces (Hilbert schemes) for rational
curves on Calabi--Yau manifolds. Our results agree with those obtained by
more traditional mathematical methods in the limited number of cases for
which the latter analysis can be carried out.  Finally, we make explicit
some striking relations between instanton corrections for various Yukawa
couplings, derived from the associativity of the operator product algebra.

\Date{12/93}

\newsec{Introduction}

\nref\rconj{
L.J. Dixon, in {\it Superstrings, Unified Theories, and
Cosmology 1987}, (G. Furlan et. al., eds.), World Scientific, 1988, p. 67\semi
W. Lerche, C. Vafa, and N.P. Warner, Nucl. Phys. {\bf B324} (1989)
427.}%
\nref\rCLS{P. Candelas, M. Lynker, and R. Schimmrigk, Nucl. Phys. {\bf B341}
(1990) 383.}%
\nref\rGP{B.R. Greene and M.R. Plesser, Nucl. Phys. {\bf B338} (1990) 15.}%
\nref\rALR{P.S. Aspinwall, C.A. L\"utken, and G.G. Ross, Phys. Lett. {\bf 241B}
(1990) 373.}%
\nref\rCDGP{P. Candelas, X.C. de la Ossa, P.S. Green, and L. Parkes,
Phys. Lett. {\bf 258B} (1991) 118;
Nucl. Phys. {\bf B359} (1991) 21.}%

Calabi--Yau threefolds were originally introduced into string theory
to provide  six compact spatial
dimensions which complement   four  Minkowski spacetime
directions to yield a consistent ten dimensional background for
string propagation. From a more general perspective, Calabi--Yau
threefolds can be target spaces for two dimensional
supersymmetric ($N = 2$) conformally
invariant nonlinear sigma models with $c = 9$---this number arising from
three times the  complex dimension of the target space. Such superconformal
field theories are of interest for a number of reasons including applications
to
string backgrounds, critical systems and issues in mathematical physics.
In the latter category, the recent conjectures \rconj, evidence from
numerical studies \rCLS,
explicit construction \rGP, and applications
\refs{\rALR, \rCDGP} of mirror symmetry are indications
of a deep mathematical structure that, at present, is best understood from
the physical viewpoint.

\nref\rDRM{D.R. Morrison, in
{\it Essays on Mirror Manifolds}, (S.-T. Yau, editor), International Press,
 1992, p. 241.}%
\nref\rFont{A. Font,
Nucl. Phys. {\bf B391} (1993) 358%
.}%
\nref\rKT{A. Klemm and S. Theisen,
Nucl. Phys. {\bf B389} (1993) 153.}%
\nref\rKatz{S. Katz,
``Rational curves on Calabi--Yau manifolds:
verifying predictions of mirror symmetry,''
in {\it Algebraic Geometry}, (E.~Ballico, editor), Marcel Dekker, to appear;
alg-geom/9301006.}%
\nref\rES{G. Ellingsrud and S.A. Str{\o}mme, {\it The number of twisted
cubic curves on the general quintic threefold}, University of Bergen report
no.~63-7-2-1992.}%
\nref\rESprivate{G. Ellingsrud and S.A. Str{\o}mme, private communication.}%
\nref\rW{E. Witten, Surveys in Diff. Geom. {\bf 1} (1991) 243.}%

The focus on $ c = 9$, as mentioned, has its origin in the string theoretic
applications of Calabi--Yau manifolds. The mathematical physics applications,
however, are of interest in the more general setting of $ c = 3 d$
corresponding to Calabi--Yau $d$-folds. It is the purpose of the present paper
to study mirror manifolds for more general values of $d$.
There are a couple of motivations for this study. First, there are some
incompletely understood aspects of mirror symmetry.
It is one of our hopes that by studying
mirror symmetry for general dimension, the special features of dimension three
can be
suppressed
and hence allow focus on the true (dimension independent)
mathematical and physical
characteristics responsible for mirror symmetry.
 Second, mirror manifolds
in dimension three have proven themselves to be a powerful calculational
tool. In particular, by making use of the mirror manifolds constructed in
\rGP, the authors of \rCDGP\ and \refs{\rDRM, \rFont, \rKT}
showed that the number of rational curves of arbitrary degree on certain
Calabi--Yau threefolds---a problem heretofore impenetrable with standard
mathematical methods---could be calculated with relative ease.
There are two special
features of dimension three in this regard.  First, rational curves
(world sheet instantons)
on a generic Calabi--Yau threefold are isolated whereas they arise in
continuous families on higher dimensional Calabi--Yau manifolds.
The analog of calculating the
number of rational curves  of a given degree
on a Calabi--Yau threefold is the calculation of properties of Chern
 classes of the parameter spaces of such curves in the higher dimensional
case.
These parameter spaces are
subspaces of the so-called {\it Hilbert schemes\/} of rational curves of a
given degree.  (These Hilbert schemes are analogous to the Grassmannian
which parameterizes rational curves of degree one.)
We will see that the integers associated with these
characteristic classes, for rational curves of arbitrary degree, are
easily calculated so long as we are in possession of the mirror of
the Calabi--Yau manifold under consideration. For the calculations associated
with degree one and degree two curves, our results have been confirmed
by more standard mathematical methods by Katz \rKatz. For degree higher
than two, however,  the latter mathematical approach
does not apply.\foot{However, very recently Ellingsrud and Str{\o}mme
have generalized their earlier work \rES\ and have verified some of our
predictions for degree three curves \rESprivate.}
A second distinction   is that
whereas there is
one type of Yukawa coupling (for each of the $(c,c)$ and $(a,c)$ rings)
on a threefold there are many more in the higher dimensional case. Each
of these couplings probes part of the chiral and antichiral primary field
ring structure and has an instanton expansion interpretable as above.
We will see that associativity of the operator product algebra gives rise
to striking relations amongst these instanton expansions. Mathematically,
these relations can likely
be established by making use of the degeneration argument
invoked by Witten in \rW.

\nref\rGPNE{B.R. Greene and M.R. Plesser, in
{\it Proceedings of the Second International {\rm PASCOS} conference},
Northeastern University, 1991.}%

The above discussion, of course, only applies to Calabi--Yau manifolds for
which we have a mirror partner. General physical reasoning lends credence
to the conjecture \rconj\ that all Calabi--Yau manifolds come in mirror pairs
(see \rGPNE\ for a review). To date, the only proven constructions
of mirror manifolds are those given in \rGP\ and hence we shall focus
on this subspace of Calabi--Yau manifolds.
For this purpose we briefly recall the main result of \rGP.

Let $W$ be a Calabi--Yau $d$-fold realized as a Fermat hypersurface
in a weighted projective space of dimension $d + 1$, $\WCP{d+1}{}$.
Let $G$ be the maximal group of diagonal scaling symmetries acting on
the homogeneous $\WCP{d+1}{}$ coordinates which preserves the holomorphic
$d$-form on $W$. Then $W$ and $M=W/G$ constitute a mirror pair.
Furthermore,
as explained in \refs{\rGP, \rGPNE}
a point crucial to the analysis of \refs{\rALR, \rCDGP} and to our study here,
is
that although the explicit arguments for constructing mirror pairs
\rGP\ are tied
to special points in moduli space (the Fermat points), deformation
arguments allow us to move away from such points via changes in either
the complex structure or the K\"ahler structure\foot{Technically it
might be difficult to establish this statement for all but local
deformations in the moduli space. We stress, however, that the results
presented in \refs{\rALR, \rCDGP} and here all rely on our deformation
reasoning
applying globally in the moduli space.}. We therefore are able to construct
families of mirror pairs by deforming from the Fermat point \rGP.

\nref\rAM{P.S. Aspinwall and D.R. Morrison, Commun. Math. Phys. {\bf 151}
(1993) 245.}%

In section II we study some aspects of the moduli space of Calabi--Yau
$d$-folds from a covariant viewpoint. Our purpose in this discussion is
not to be complete, but rather to indicate the ways in which the moduli
spaces for higher dimensional Calabi--Yau manifolds differ from the three
dimensional case.
In particular we note that these moduli spaces
have properties which differ in detail from the three dimensional case but
retain certain important qualitative features.
We give procedures which allow for the derivation of the Picard--Fuchs
equations governing the behavior of the period maps which involve complications
that are not present in the well studied case of $d = 3$. In section III
we calculate the generalized ``Yukawa couplings'' (higher point functions)
which naturally arise in this analysis
(for a variety of examples) and then
apply the methods of \rDRM\  to derive  mirror maps and hence an instanton
expansion.  These higher point functions will factor into (sums of) products
of three-point functions.
We show that the associativity of the operator product
expansion gives rise to relations amongst these three-point functions
(the ``conformal bootstrap equations'') which translate into striking
implications for the associated instanton expansions.
In section IV we rephrase
the analysis of section II in a form better suited to the incorporation
of mirror symmetry and explicit calculations.
This approach naturally yields the fundamental Yukawa couplings
(three-point functions) which we relate to the calculations in the previous
section.
In section V we give the mathematical
interpretation of the instanton expansions found in section IV
(which is most easily done in the language of topological field theory).
Full justification of the interpretation requires a topological field theory
argument along the lines of \rAM\ which is presented in an appendix.
In section VI we state our conclusions.

\newsec{Calabi--Yau Moduli Spaces for $d \gt 3$}

\nref\rE{
S. Ferrara and A. Strominger, in {\it Strings '89} (R. Arnowitt et al., eds.),
World Scientific, Singapore, 1989\semi
S. Cecotti, S. Ferrara and L. Girardello, Int. J. Mod. Phys. {\bf 4}
(1989) 2475; Phys. Lett. {\bf B213} (1988) 443.}%
\nref\rStrom{A. Strominger, Commun. Math. Phys. {\bf 133} (1990) 163.}%
\nref\rDKL{L. Dixon, V. Kaplunovsky and J. Louis, Nucl. Phys. {\bf B329}
(1990) 27.}%
\nref\rCd{P. Candelas and X. de la Ossa, Nucl. Phys. {\bf B355} (1991) 455.}%

Work over the last few years \refs{\rE,\rStrom,\rDKL,\rCd}
has established that the moduli
spaces for Calabi--Yau threefolds are {\it special K\"ahler\/} manifolds.
We recall a few characteristics of these. Special K\"ahler manifolds
are K\"ahler manifolds of restricted type, meaning that the K\"ahler
class of the manifold
 $\cal M$ is an integral class, hence it is the first Chern class
of some line bundle which we shall denote by $\cal L$. Special
geometry asserts the existence of a set of coordinates on $\cal M$ and a
gauge choice on $\cal L$ such that the K\"ahler potential $K$ is given by
\eqn\eKA{
e^{-K} =  -i(z^a {{{ \del {\cal \overline G} }} \over
{{\del \overline z^a}}}  - \overline z^a{ {{ \del {\cal  G} }} \over
{{\del z^a}}})
}
with $\cal G$  a holomorphic function of the local complex
coordinates $z^a$. As discussed in \rCDGP, this amounts to the statement
that the K\"ahler potential has a holomorphic prepotential.

One can also think of special geometry as providing an additional
restriction on the Riemann tensor of the moduli space  beyond
those implied by K\"ahlerity. This more covariant formulation
requires the existence of holomorphic sections $\kappa_{\alpha\beta\gamma}$
of ${\cal L}^2\otimes {\rm Sym} {T^*({\cal M})}^{\otimes 3}$ such that
\eqn\eCon{
R_{\overline \delta \alpha \overline \beta \gamma} =
G_{\alpha \overline \beta} G_{\gamma \overline \delta} +
G_{\alpha \overline \delta} G_{\gamma \overline \beta}  -
e^{2K} G^{\epsilon \overline \epsilon}
\kappa_{\alpha \gamma \epsilon} {\overline \kappa}_
{\overline \beta \overline \delta \overline \epsilon}
}
where $G_{\alpha \overline \beta}$ is the K\"ahler metric on $\cal M$.
It follows that
\eqn\eYukk{
 \kappa_{\alpha \gamma \epsilon} = \del_{\alpha} \del_{\gamma}
\del_{\epsilon} {\cal G} \ .
}

When $\cal M$ is the moduli space of complex structures on a Calabi--Yau
threefold the structures of special geometry are realized as follows
\refs{\rStrom,\rCd}.
A section of
$\cal L$ is a choice of a holomorphic $3$-form $\Omega(z)$ on the
Calabi--Yau space corresponding to the point $z\in \cal M$; the K\"ahler
potential is
\eqn\eklr{
e^{-K} = \int \Omega \wedge \overline{\Omega}\ ;
}
the sections $\kappa_{\alpha\beta\gamma}$ are the Yukawa couplings
and may be written
\eqn\eYuk{
\kappa_{\alpha \gamma \epsilon}(z) =
\int_{M_z} \Omega \wedge \del_{\alpha} \del_{\gamma}
\del_{\epsilon} \Omega \ .
}
The essential point
is that the Yukawa couplings and the K\"ahler potential on $\cal M$
are both determined by the single holomorphic function $\cal G$,
and they {\it depend holomorphically on parameters}.  (This point will be
important later.)  There is a similar
structure on the K\"ahler moduli space \rDKL.

\nref\rsugra{B. de Wit and A. Van Proeyen, Nucl. Phys. {\bf B245}
(1985) 89\semi
B. de Wit, P. Lauwers and A. Van Proeyen, Nucl. Phys. {\bf B255}
(1985) 569\semi
E. Cremmer, C. Kounnas, A. Van Proeyen, J.P. Derendinger, B. de Wit and
L. Girardello, Nucl. Phys. {\bf B250} (1985) 385.}%

Special geometry was first defined as a consistency requirement
arising in the study of $N=2$ supergravity \rsugra. String theory
associates $N=2$ supergravity models to Calabi--Yau threefolds; it thus
followed that moduli spaces of Calabi--Yau threefolds must exhibit this
structure. When we discuss Calabi--Yau manifolds of dimension larger
than three string theory leads to no such association and indeed, as
we shall see, moduli spaces will not be special K\"ahler. In heuristic terms,
whereas special geometry implies that the K\"ahler potential and Yukawa
couplings are determined by a single holomorphic function, a number of
holomorphic functions (associated with the independent kinds of Yukawa
couplings) determine these features in the higher dimensional setting.

\subsec{Mathematical Preliminaries}

\nref\rTian{G. Tian, in
{\it Mathematical Aspects of String Theory}, (S.-T. Yau, editor),
World Scientific, 1987, p. 629.}%
\nref\rFerrar{A. Ceresole, R. D'Auria, S. Ferrara, W. Lerche, and J. Louis,
Int. J. Mod. Phys. {\bf A8} (1993) 79.}%

We now turn to a study of the moduli space $\cal M$ of complex structures
on a Calabi--Yau $d$-fold $M$. We denote by $\Omega (z)$ a chosen
holomorphic $d$-form
on the Calabi--Yau space corresponding to the point $z$ in $\cal M$.
As in \refs{\rStrom,\rTian},
$\Omega$ is naturally thought of as a section of the Hodge bundle
${\cal H}$ over $\cal M$ (the fibers of which are $H^d(M,\IC)$).
As the parameters $z$ vary, $\Omega(z)$ spans a
holomorphic line bundle $\cal L\subset \cal H$ whose first Chern
class is the K\"ahler form on $\cal M$.%
\foot{The bundle ${\cal H}\otimes {\cal L}^{-1}$
has fibers which can be canonically identified with $H^0(M,\Lambda^0T)
\oplus H^1(M,\Lambda^1T)\oplus\cdots\oplus H^d(M,\Lambda^dT)$.
As pointed out by Strominger \rStrom, these fibers can also be identified
with $H^d(M,\IC)$; however, {\it that\/} identification is {\it not\/}
canonical.  Our treatment thus differs from \rStrom\ in considering
$\Omega$ as a section of $\cal H$ and not
${\cal H}\otimes{\cal L}$.}
We derive differential equations
for the periods of $\Omega(z)$ (over a suitable family of cycles to be
discussed), called  Picard--Fuchs equations, in a manner similar in spirit
to \refs{\rStrom, \rFerrar}.

To do so, we make use of the fact
that if $s$ is a $(p,q)$ form valued section of $\cal H $,
its covariant derivative contains $(p-1, q+1)$ valued pieces. By
beginning with $\Omega$, taking successive derivatives, and
isolating
the appropriate piece, we can construct a sequence of  maps
 of the form%
\foot{We shall see later that this procedure corresponds
to generating a partial basis for the chiral ring of the associated
conformal field theory by successive operator products of the
marginal fields.}
\eqn\eSeq{
(d,0) \rightarrow (d-1,1) \rightarrow \cdots  \rightarrow
(1,d-1) \rightarrow (0,d) \rightarrow 0\ .
}
In \rFerrar\ this sequence was used to generate the Picard--Fuchs
equation for a Calabi--Yau threefold, and we will find that for
one-parameter families we can extend this result to $d>3$.

The first step in \eSeq , as shown in \rStrom , is accomplished with
the aid of the covariant derivative $D =\nabla+\omega$, where $\nabla$
is the flat metric-compatible connection\foot{As is common in the modern
mathematical literature, we use the terms ``connection'' and
``covariant derivative'' more or less interchangeably.}
 on $\cal H$, and $\omega$ is a
correction term characterized by the property that $D$ acts covariantly on
sections of ${\cal L}$.  The components of the one-form $D\Omega$ span
$H^{(d-1,1)}(M)$, providing the first map required in \eSeq.
The connection $\nabla$ will be discussed in greater
detail in section IV; here we list the following  properties
which we will use.

(1)
$\nabla$ is a flat holomorphic connection compatible with the
metric on $\cal H$ given by the norm
\eqn\enorm{
\parallel \eta\parallel = i^{d^2} \int_M \eta\wedge\bar\eta\ .
}

(2) $\nabla$ maps $\cal H$ holomorphically to ${\cal H}\otimes T^*({\cal M})$;
in its action on $\cal H$ it is constrained by the property that it
maps  ${\cal F}^p $ to $ {\cal F}^{p-1} \otimes T^*({\cal M})$, where
${\cal F}^p = \oplus_{p' \ge p} H^{p',d - p'}$. (This property is known as
Griffiths transversality).

(3) The correction term $\omega$ can be computed
using the chosen section $\Omega$ as
\eqn\eomega{
\omega=-\partial\log\langle \overline{\Omega}|\Omega\rangle \ .
}

Since ${\cal L}$ is a line bundle, the covariant derivative
$D=\nabla+\omega$
admits another interpretation---it essentially
coincides with the metric connection
on the tensor product bundle ${\cal H}\otimes{\cal L}^{-1}$.
This is seen as follows.  The metric connection $D_{\cal L}$
on ${\cal L}$ has the property
\eqn\eDL{D_{\cal L}(f\Omega)=df\,\Omega-f\omega\,\Omega}
with $\omega$ as in \eomega; the dual bundle ${\cal L}^{-1}$ will have a
connection ${\cal D}_{{\cal L}^{-1}}$ satisfying
\eqn\eDLinv{{\cal D}_{{\cal L}^{-1}}(f\Omega^{-1})
=df\,\Omega^{-1}+f\omega\,\Omega^{-1}.}
If we write a section
of ${\cal H}\otimes{\cal L}^{-1}$
in the form $\alpha\otimes\Omega^{-1}$ (with $\alpha$ a section of
${\cal H}$) then we find
\eqn\eDH{\eqalign{{\cal D}_{{\cal H}\otimes{\cal L}^{-1}}
(\alpha\otimes\Omega^{-1})&=
{\cal D}_{\cal H}(\alpha)\otimes\Omega^{-1}
+\alpha\otimes{\cal D}_{{\cal L}^{-1}}(\Omega^{-1}) \cr
&=\nabla(\alpha)\otimes\Omega^{-1}+\omega\alpha\otimes\Omega^{-1}
=D(\alpha)\otimes\Omega^{-1}}}
so that ${\cal D}_{{\cal H}\otimes {\cal L}^{-1}}$
is calculated in terms of the covariant
derivative $D$.
It is thus natural to think of the sequence of maps in \eSeq\ as
constructed in ${\cal H}\otimes{\cal L}^{-1}$, and this is the
route we will follow in the next subsection.

The subbundle ${\cal L}\subset{\cal H}$ whose fibers
are $H^{d,0}(M)$ gives rise (upon tensoring with ${\cal L}^{-1}$)
to a subbundle ${\cal O}\subset{\cal H}\otimes{\cal L}^{-1}$
isomorphic to the trivial bundle,
whose fibers are $H^0(M,\Lambda^0T)$.  This subbundle comes equipped
with a completely canonical section $\boldone$ (the constant function
$1$ on the moduli space).  To calculate
${\cal D}_{{\cal H}\otimes{\cal L}^{-1}}(\boldone)$,
we must write $\boldone=\Omega\otimes\Omega^{-1}$, and we then find
\eqn\eDHone{{\cal D}_{{\cal H}\otimes{\cal L}^{-1}}(\boldone)=
D(\Omega)\otimes\Omega^{-1}=(\nabla(\Omega)+\omega\,\Omega)\otimes\Omega^{-1}.}
This is independent of the choice of $\Omega$.

Notice that in this description the {\it a priori\/} arbitrariness in the
choice
of $\Omega$ becomes irrelevant. This description thus gives a nice
resolution to an
uncomfortable apparent asymmetry in one aspect of mirror
manifolds, as we now briefly
mention.

Often times, in the discussion of mirror manifolds, the symmetry between
the ``vertical'' cohomology    $\oplus_p H_{\tilde M}^{p,p}$ on $\tilde M$
and the ``horizontal'' cohomology $\oplus_p H_{M}^{d-p,p}$ on
its mirror $M$ has been emphasized. On closer inspection, though,
there lurks an uncomfortable asymmetry between these two structures:
there is a canonical section of $H_{\tilde M}^{0,0}$ which naturally enters
the discussion---namely the constant section $\boldone$. On the other hand,
there is no canonical choice of section of $H_{M}^{d,0}$
 which is the mirror
cohomology group of $H_{\tilde M}^{0,0}$.
The resolution of this apparent asymmetry which we present here is based on
two observations. First, at a more fundamental level, mirror manifolds
respect a symmetry which exchanges $\oplus_p H^p(M, \Lambda^p T^*)$
with $\oplus_p H^p(\tilde M, \Lambda^p T)$.
The former is canonically isomorphic to $\oplus_p H_{\tilde M}^{p,p}$
while the latter
as we have discussed,
is canonically isomorphic to ${\cal H} \otimes {\cal  L}^{-1}$. For $p = 0$
each of these does have a canonical section, namely, $\boldone$.
Second, as we have just calculated, the covariant derivative which
we use to realize \eSeq\ is also independent of the choice of $\Omega$.
Hence, in this description everything is manifestly mirror symmetric.

\subsec{Picard--Fuchs Equations}

Following the basic strategy outlined above, we now consider successive
derivatives of $\boldone$, attempting at each stage to isolate the component
of pure type $(p,q)$ for the smallest possible value of $p$.%
\foot{Note that we are using ``$(p,q)$'' a bit loosely
here; we should actually shift everything to sections
of ${\cal H}\otimes{\cal L}^{-1}$, where a $(p,q)$ form will turn into
a section of $H^q(M,\Lambda^qT)$.}
An invariant way of describing this procedure is to introduce an
additional correction term $\omega_p$ with the properties that
(1) ${\cal D}_{{\cal H}\otimes{\cal L}^{-1}}+\omega_p$
acts covariantly  on sections of ${\cal F}^p\otimes{\cal L}^{-1}$, and (2)
${\cal D}_{{\cal H}\otimes{\cal L}^{-1}}+\omega_p$ maps
${\cal F}^p\otimes{\cal L}^{-1}$ to $({\cal F}^p\otimes{\cal L}^{-1})^\perp$.
By Griffiths transversality, it follows that
$({\cal D}_{{\cal H}\otimes{\cal L}^{-1}}+\omega_p)(\chi)$
must precisely pick out the
$(p-1,q+1)$ piece of ${\cal D}_{{\cal H}\otimes{\cal L}^{-1}}(\chi)$.

Above we observed that in the first step of \eSeq\ we in fact produce
a basis for the cohomology group $H^{(d-1,1)}$. In subsequent steps
this nice property will no longer hold (as pointed out in \rStrom).
The components of successive derivatives of $\Omega$ will span a
subbundle of ${\cal H}$ that we will term the {\it primary
horizontal subspace}. This is most easily understood in
${\cal H}\otimes{\cal L}^{-1}$ where it comprises the subspace
generated by successive cup products of the
elements of $H^1(T({\cal M}))$; we will restrict our
attention to this subspace. Thus, to verify
at each step that we have correctly projected to
$({\cal F}^p\otimes{\cal L}^{-1})^\perp$ we will check that components
along the elements of a basis for the primary subspace constructed in
previous steps vanish. The metric \enorm\ restricts (by the Hodge--Riemann
bilinear relations) to a nondegenerate pairing on the primary
subspace, so we will use this to compute these projections.

To avoid cluttering the notation, we will drop the subscripts and let
$D$ denote the metric connection on
${\cal H}\otimes{\cal L}^{-1}\otimes\Lambda^n T^*({\cal M})$ (with the
appropriate value of $n$ determined by context) derived from
${\cal D}_{{\cal H}\otimes{\cal L}^{-1}}$ by adding an appropriate number
of Christoffel connection terms.
The first step in \eSeq\ is realized quite simply as
$X^{(1)} = D \boldone$; let us show that this is of pure type.
To determine the $(d,0)$ component we use the metric compatibility to
evaluate
\eqn\eProja{
\langle \overline\boldone | X^{(1)} \rangle =
\partial\langle\overline\boldone | \boldone\rangle -
\langle  D \overline \boldone | \boldone\rangle = 0
\ .}
The inner product $\langle\ | \ \rangle$
on ${\cal H}\otimes{\cal L}^{-1}$ is the one derived from \enorm;
\eProja\ thus equates one-forms.
The first term is trivially zero. To see that the second term
vanishes, recall that $\boldone$ is a holomorphic section of a
holomorphic bundle on which $D$ is a holomorphic connection. Thus
writing $X^{(1)} = \chi^{(1)}_\alpha\otimes\Omega^{-1} dz^{\alpha}$
we have seen that $\chi^{(1)}_\alpha$ is purely of type $(d-1,1)$.
In fact, of course, the components of $X^{(1)}$ span $H^1(T)$, realizing
the Kodaira--Spencer isomorphism
 between this cohomology group and the fibers of $T({\cal M})$.

At the next stage we again set $X^{(2)} = D X^{(1)}$.
A computation essentially identical to \eProja\ shows that $X^{(2)}$
has no $(d,0)$ component. To
determine the $(d-1,1)$ component we compute the inner products with
our basis
\eqn\eProjb{
\langle \overline{X^{(1)}_\alpha} | X^{(2)}_{\beta\gamma}\rangle =
{\cal D}_\beta\langle \overline{X^{(1)}_\alpha} | X^{(1)}_\gamma\rangle
- \langle D_\beta D_{\overline\alpha}\overline{\boldone} |
X^{(1)}_\gamma\rangle
\ .}
In the first term, $\cal D$ is the metric
connection\foot{This connection is {\it not}\/ holomorphic. For further
details on manipulating connections such as this one, see Appendix A.}
 on
$T^*({\cal M})\otimes\overline{T^*}({\cal M})$ and the tensor on which
it acts is calculated to be
the K\"ahler metric yielding zero. In the second term we can use
once more the holomorphicity of $\boldone$ to obtain a commutator term
\eqn\eProjc{
\langle [D_\beta,D_{\overline\alpha}]\,
\overline{\boldone}|X^{(1)}_\gamma\rangle
\ .}
This commutator of covariant derivatives is just the curvature of the
bundle in which they act---in this case because $\cal H$ is flat we
obtain the curvature of ${\cal L}^{-1}$. Since this does not change
type, the previous calculation shows that this term also vanishes.
Hence, in the sense used above, $X^{(2)}$ is of pure type
$(d-2,2)$.

We note here that for $d=3$, the above steps suffice
\refs{\rStrom,\rFerrar} to compute the differential equation. The
reason is that in this case we already have in hand a basis for
$H^{(1,2)}$, given by the complex conjugate of our basis for
$H^{(2,1)}$. Expressing $X^{(2)}$ in terms of $\overline{X^{(1)}}$ we
can complete \eSeq\ by considerations similar to those above. This, of
course, will not suffice for $d>3$. This does point to one difficulty
we will encounter, however. The components of the forms generated in
realizing \eSeq\ will by definition span the primary subspace, but in
general they will not be linearly independent.

We now attempt, therefore, to continue by setting
$X^{(3)} = D X^{(2)}$.
The above reasoning ensures that besides
the desired $(d-3,3)$ part, the only other possible component is of type
$(d-2,2)$. To probe for the latter
we compute
\eqn\eProjd{\langle \overline {X^{(2)}_{\alpha\beta}}  |
X^{(3)}_{\gamma\delta\epsilon} \rangle }
which equals
\eqn\eProje{
{\cal D}_\epsilon
\langle \overline {X^{(2)}_{\alpha\beta}} | X^{(2)}_{\gamma\delta} \rangle  -
\langle D_\epsilon\overline{X^{(2)}_{\alpha\beta}}  |
 X^{(2)}_{\gamma\delta} \rangle
\ .}
Let us deal with the second term in \eProje\ first. Quite generally,
any expression of the form
\eqn\eProjgen{
\langle D_\alpha D_{\overline \alpha_1} \ldots
D_{\overline \alpha_k} \overline\boldone
 \vbar \tilde X^{(k-1)}_{\alpha_1\ldots\alpha_{k-1}}\rangle
}
vanishes identically if $\tilde X^{(k-1)}$ is of pure type
$(d-k+1,k-1)$. To prove this, we commute the $D_\alpha$  successively to
the right. Examination of each of the resulting terms shows that they are all
of at most type $(k-2, d-k+2)$ and hence yield zero inner product. Using this
result we see that
\eqn\eProjf{
\langle \overline {X^{(2)}_{\alpha\beta}}  |
X^{(3)}_{\gamma\delta\epsilon} \rangle =
{\cal D}_\epsilon\langle \overline {X^{(2)}_{\alpha\beta}} |
X^{(2)}_{\gamma\delta} \rangle
\ .}
Now, this expression is generally nonzero and hence $X^{(3)}$ is not
of pure type $(d-3,3)$.  We can, however, seek to write
\eqn\eFacout{
\tilde X^{(2)}_{\alpha\beta} = S^{\gamma\delta}_{\alpha\beta}
X^{(2)}_{\gamma\delta}
}
with $S$ so chosen that $D \tilde X^{(2)}$ is of pure type $(d-3,3)$.%
\foot{This can only determine $S$ up to covariantly constant factors.}
In general it is not clear how to construct such an $S$, since this
requires untangling the linear dependence of the components of $X^{(2)}$
mentioned above. (In the case
$d=3$ it was found \rFerrar\  using the expression for
$X^{(2)}$ in terms of $\overline{ X^{(1)}}$.) However, in the
particular case of interest to us here, for a one-parameter family, it
is not difficult to proceed; we restrict attention to this case now.
The Greek indices all take but one value, and will occasionally be
omitted. Then the simple solution is $S =
(\langle\overline{X^{(2)}}|X^{(2)}\rangle)^{-1}$.

One can imagine continuing in this fashion as follows. Let
$X^{(k)} = D^k \boldone$, then this will be at most of type
$(d-k,k)$. Reasoning as above, we will need to define
$X^{(k)} = D \tilde X^{(k-1)} = D(S^{(k-1)} X^{(k-1)})$ with $S$ chosen so that
$\langle \overline{X^{(k-1)}} | \tilde X^{(k-1)}\rangle$ is covariantly
constant. Then the arguments of the preceding paragraphs will show that
$X^{(k)}$ is of pure type. The solution for $S$ given above extends as
\eqn\eSS{
S^{(k)} = {\left( \langle \overline{X^{(k)}} | X^{(k)}\rangle\right)}^{-1}
\ .}
This procedure thus realizes the sequence \eSeq\ and terminates
of course at the $(d+1)$-st step, yielding a differential equation of this
order for $\boldone$ or equivalently for $\Omega$ and hence its
periods. This is the Picard--Fuchs equation, in a form very similar to
that obtained in \rFerrar\ for $d=3$:
\eqn\ePF{
D(S^{(d-1)}D)\ldots(S^{(3)}D)S^{(2)}DD\boldone=0
\ .}

\subsec{Analogs of Special Geometry}

As mentioned earlier, for $d=3$ it is well known that Calabi--Yau moduli
spaces are special K\"ahler manifolds which, for example, can be
characterized by \eCon. The moduli spaces for Calabi--Yau manifolds for
$d \gt 3$ do not satisfy \eCon\ but they do respect particular constraints on
their respective Riemann tensors as we now briefly indicate.

\nref\rBCOV{M. Bershadsky, S. Cecotti, H. Ooguri, C. Vafa,
{\it Kodaira--Spencer theory of gravity and exact results for quantum
   string amplitudes}, Harvard preprint HUTP-93-A025, hep-th/9309140.
}

For arbitrary $d$ we have
\eqn\eCurv{
[D_\alpha,D_{\overline \beta}]\, X^{(1)}_\gamma =
-G_{\alpha \overline \beta} X^{(1)}_\gamma
+ R_{\alpha \overline{\beta}  \gamma}^\delta X^{(1)}_\delta
}
simply expressing the curvature of $D$ as it acts on this bundle and
recalling that $\nabla$ is a flat connection on $\cal H$.
Solving for the Riemann tensor we thus have
\eqn\eRiem{
R_{\alpha \overline{\beta} \gamma \overline \delta } =
G_{\alpha \overline \beta} G_{\gamma \overline \delta } +
G_{\alpha \overline \delta} G_{\gamma \overline \beta}
+ e^K \int \chi^{(1)}_{\overline \delta} D_{\overline \beta }
D_\alpha \chi^{(1)}_\gamma
\ .}
The last term on the right-hand side would, in the notation of the
previous subsection, be written after integration by parts as
\eqn\eRb{
\langle \overline{X^{(2)}_{\beta\delta}} | X^{(2)}_{\alpha\gamma}\rangle
\ ,}
in which form this equation recently appeared in \rBCOV.

We wish to find a constraint on the Riemann tensor which is
written explicitly in terms of the higher dimensional analog of
\eYuk. This requires an explicit
evaluation of the integral on the right hand side of \eCurv. In the case
of $d= 3$ this is easy to do since $D_\alpha \chi^{(1)}_\beta$ is pure
type $(1,2)$ and
is readily expressed in terms of $\chi^{(1)}_{\overline \beta}$
(and similarly for
the complex conjugate situation which also arises in \eCurv).
When the tensor $S^{(2)}$ (as described above) exists, this can
be explicitly carried out in a similar manner for $d = 4$ and leads, after some
algebra, to
\eqn\eCurvb{
R_{\alpha \overline{\beta} \gamma \overline \delta } =
G_{\alpha \overline \beta} G_{\gamma \overline \delta } +
G_{\alpha \overline \delta} G_{\gamma \overline \beta}
+ e^K B^{\alpha'\gamma'\overline{\beta}'\overline{\delta}'}
\kappa_{\alpha'\gamma'\alpha\gamma}
\overline{\kappa}_{\overline{\beta}'\overline{\delta}'
\overline{\beta}\overline{\delta}}
}
where $B = S^{(2)}$ and $\kappa$ is the Yukawa coupling defined in any
dimension $d$
by
\eqn\ekappadef{
\kappa_{\alpha_1\ldots\alpha_d} = \int \Omega \wedge \del_{\alpha_1}\ldots
\del_{\alpha_d} \Omega
\ .}
Written in this way we see the similarity to $d = 3$, the main difference
being the tensor $B$ (which is essentially the inverse of $\langle
\overline{\chi^{(2)}_{\beta\delta} } \vbar  \chi^{(2)}_{\alpha\gamma}
 \rangle $)
taking the place of ${G}^{\alpha\overline\beta}$
(which arises from the inverse of $\langle \overline\chi^{(1)}_\beta \vbar
\chi^{(1)}_\alpha\rangle $).

In the case of a one-parameter family where the tensors $S^{(k)}$ exist
and the analysis above is  valid,
we can explicitly compute (again omitting the indices)
\eqn\eB{
B = e^K/(2 G^2 - R)
\ .}
Thus
\eqn\eCurvbb{
(R - 2 G^2)^2 = e^{2K} \kappa \overline \kappa
\ .}
The Hodge--Riemann bilinear identities ensure that $2 G^2 - R$
is positive
and hence  we find (replacing the index placeholders)
\eqn\eCurvc{ R_{\alpha\overline\alpha\alpha\overline\alpha} =
 2 G_{\alpha\overline\alpha}^2 -
 e^K  | \kappa_{\alpha\alpha\alpha\alpha}|
\ .}

The approach can be pursued further. This can be done
by explicitly evaluating the right hand side of \eRiem. Alternatively
(and somewhat easier to calculate) one can pursue the direct analog of
the three (or four) dimensional calculation and consider
$[ D_{\alpha}, D_{\overline \beta} ] \,X^{(j)}$ with
$j = [d - 1]/2$ as before. Since $X^{(j)}$ is a section of
${\cal H} \otimes{\cal L}^{-1}\otimes T^*({\cal M})^j$ the commutator
involves sums of terms involving the Riemann tensor and the
metric on moduli space. The advantage of operating on a section of this
particular bundle is that for this value of $j$ the action of $D_{\alpha}$
pushes us over the ``half-way'' point, thus allowing us to reexpress the
result in terms of the complex conjugate basis (as discussed earlier).
This
facilitates direct calculation of the constraint on the Riemann tensor.
For example, in the case of a one-parameter family with $d = 5$ we find
\eqn\edequalsfive{
({R_{;\alpha}  \over 2 {G}^2 - R})_{; \overline
 \alpha} + {|\kappa|^2  \over  (2 {G}^2 - R)^2} \, e^{2K} =
3( {G} -  {G}^{-1} R)
}
where $R$ is the Riemann tensor and $G$  is the metric on moduli space.
In general when one attempts to evaluate the right-hand side of
\eRiem\ in terms of the Yukawa couplings the expressions become
complicated for large $d$.

\newsec{Yukawa Couplings,  Series Expansions and Factorization}

In the previous section we described the general structure of moduli
spaces for Calabi--Yau manifolds in general dimension $d$. Our aim is
to apply mirror symmetry to these manifolds, and to this end we will
in this section introduce the physical theories related to the
geometrical constructs. We will then compute the correlation functions
of marginal chiral primary operators in a set of models and exhibit
the series expansions predicted for these functions by mirror symmetry.
Finally, we will show how these functions are predicted to factorize
in terms of more fundamental correlators and extract some highly
nontrivial predictions regarding this factorization. Computing the
fundamental couplings will require the introduction of some additional
structure and this will be the subject of the next section.

\nref\rWitten{E. Witten, in
{\it Essays on Mirror Manifolds}, (S.-T. Yau, editor), International Press,
 1992, p. 120.}%

Given a Calabi--Yau space $M$ equipped with a complex structure,
K\"ahler metric and $B$-field, we can define two
different topological field theories. The description of the previous
section is well-suited to a discussion of
the {\bf B} model (in the terminology of
\rWitten ). In this theory the observables are naturally described by
the space
\eqn\eBHS{
H_B = {}\oplus_{p,q=0}^d H^p(\Lambda^q T)
}
where $T$ is the holomorphic tangent bundle to $M$. The correlation
functions of the model are computable exactly in terms of geometrical
quantities. Given ${\cal O}_i \in H^{p_i}(\Lambda^{q_i} T)$ the
correlation function vanishes unless $\sum  p_i = \sum q_i = d$ and
when nonzero is given by
\eqn\eBcor{
\langle {\cal O}_1 \cdots {\cal O}_s\rangle =
\int_M \Omega_{i_1\ldots i_d}{\cal O}_1\wedge
\cdots\wedge{\cal O}_s\wedge\Omega,
}
where the notation means tangent indices are contracted with $\Omega$
and the forms cupped together. Deformations of
the complex structure of $M$ are related of course to the observables
corresponding to $H^1(T)$. (These are the marginal operators).
The nonzero correlators of these can be rewritten as
\eqn\eNP{
\eqalign{
\langle X^{(1)}_{\alpha_1}\cdots X^{(1)}_{\alpha_d}\rangle &\equiv
\kappa_{\alpha_1\dots\alpha_d} \cr
&= \int_M \Omega \wedge D_{\alpha_1} \dots D_{\alpha_d} \Omega \ ,\cr
}}
a quantity which was seen to play a role in the discussion of
section II.

\nref{\rDistler}{J. Distler,
{\it Notes on $N{=}2$ sigma models}, Princeton preprint PUPT-1365,
hep-th/9212062.}

There is a second topological field theory associated to a Calabi--Yau
manifold $M$,
the {\bf A} model of \rWitten.
For clarity below we rename the Calabi--Yau manifold to $\tilde M$, but
we stress that it is {\it not}\/ necessary to change the manifold in
order to define the {\bf A} model.
In this theory the observables
naturally correspond to the de Rham cohomology of $\tilde M$. The
parameter space of this model is a complexification of the K\"ahler
cone of $\tilde M$; all relevant quatities are completely invariant under
variations of the complex structure. The correlation functions in the
{\bf A} model are defined as sums over homotopy classes of maps from
the worldsheet (which we take in all cases to be simply \CP1, other
topologies have recently been considered in \rBCOV) to $\tilde M$. In each
class the contribution may be shown to localize on holomorphic maps,
and the contribution of each such ``instanton sector'' is weighted by
the exponential of the pullback of the K\"ahler form of $\tilde M$ (evaluated
on the fundamental class of the worldsheet). These series are
expected to have a
finite radius of convergence about a
``large radius limit'' point deep in the interior of the K\"ahler
cone; the leading term is the intersection matrix of $\tilde M$.
The nonzero correlators of marginal operators
$\tilde X_\alpha^{(1)}$
can be written
\eqn\eNPM{
\eqalign{
\langle \tilde X^{(1)}_{\alpha_1}\cdots \tilde X^{(1)}_{\alpha_d}\rangle
  &\equiv
{\tilde\kappa}_{\alpha_1\dots\alpha_d} \cr
&=
\int_{\tilde M} \tilde \Omega^{a_1\dots a_d}
A_{a_1 \overline a_1} \dots A_{a_d \overline a_d} \overline{
\tilde \Omega}{}^{\overline a_1
\dots \overline a_d}  + {\rm instanton\ corrections} }}
where $\tilde \Omega$ is a completely antisymmetric tensor field
needed to normalize the topological correlation functions.
In familiar applications a particularly natural choice for
the latter data on $\tilde M$ has been made: namely,
a completely antisymmetric
tensor field which is
constant on the K\"ahler moduli space. Although not usually
emphasized, we point out that, although natural, this {\it is\/} a choice and
mirror symmetry predicts that there is a corresponding choice for the
data on $M$ such that \eNP\ and \eNPM\ are equal.\foot{A similar
observation has been made independently by Distler \rDistler.}

The instanton
contributions to \eNPM\ from nontrivial sectors are related (as we discuss in
detail in section V) to certain characteristic classes of the
moduli space of holomorphic maps of the appropriate homotopy type,
and the extraction of explicit results on these has been one of the
most successful applications of mirror symmetry.
This application is based upon the following fact: If $M$ and $\tilde
M$ are mirror manifolds, then the {\bf A} model constructed from $\tilde M$
is isomorphic as a topological field theory to the {\bf B} model
constructed from $M$. In more detail, this means that mirror
symmetry implies the existence of a ``mirror map'' from the
complexified K\"ahler cone of $\tilde M$ to the moduli space of complex
structures on $M$, and at each point a mapping of the spaces of
observables in the two models, such that these maps preserve the
correlation functions of the topological field theory.%
\foot{In fact, as is well known, this statement is weaker than the
strongest one implied by mirror symmetry---which implies in fact an
isomorphism of the superconformal $\sigma$ models based upon $M$ and
$\tilde M$, but this version is sufficient for all of our applications
here.} In practice, to study the
properties of rational curves on a manifold $\tilde M$ one constructs the
mirror manifold $M$ and computes the {\bf B} model correlation
functions as we will do below. One then finds the location in moduli
space of the ``large complex structure limit'' point (mirror to the
large radius point), about which one expands the correlators. To
interpret the coefficients of the expansion one must expand in
coordinates related by the mirror map to the coefficients of the
K\"ahler form on $\tilde M$ in terms of a fixed basis for $H^2(\tilde M)$.
We will
find these coordinates using an {\it ansatz\/} for the mirror map, first
proposed in \rCDGP\ and recently explained in \rBCOV. These points
will be discussed in more detail in the sequel.

\subsec{The Computation}

\nref\rRoan{S.-S. Roan, Int. Jour. of Math. {\bf 2} (1991) 439.}%

We now present a class of examples for which we perform the
computations explicitly. All of these will be one-parameter families of
Calabi--Yau manifolds (i.e.\ $h^{d-1,1}(M) = 1$). In particular, for
simplicity, we will
consider families $M_{\psi}^{(d)}$ of Calabi--Yau hypersurfaces
constructed as follows. Let $W_{\psi}^{(d)}$ be a hypersurface in
$\CP{d+1}$ determined in terms of homogeneous coordinates $z_i$
by the equation%
\foot {We note that one could easily extend our analysis to include
cases involving weighted projective spaces (which even for the case of
hypersurfaces with $h^{1,1} = 1$ become quite numerous with increasing
$d$).}
\eqn\eFamily{
P(z;\psi) = z_1^{d+2} + \cdots + z_{d+2}^{d+2} - (d+2) \,\psi\,
z_1 z_2 \cdots z_{d+2} = 0
\ .}
This defines a family of Calabi--Yau manifolds with $h^{1,1}(W) = 1$.
Define $M_{\psi}^{(d)}$ by the quotient construction
\eqn\eMirr{
M_{\psi}^{(d)} = W_{\psi}^{(d)} / {(\BZ_{d+2})^d}
\ .}
By the arguments of \rGP, the family $M_{\psi}^{(d)}$ lies in the
``mirror'' parameter space, and indeed one verifies that $h^{d-1,1}(M) = 1$
\rRoan.
The parameter $\psi$ is a coordinate on the space of complex structures
on $M_{\psi}^{(d)}$. In terms of the mirror manifold
$\tilde M_{\psi}^{(d)}$ it serves as a coordinate on the complexified
K\"ahler cone. Note that $\tilde M_{\psi}^{(d)}$ is a deformation
of $W_{\psi}^{(d)}$, so
computations on $M$ will yield information about rational curves on $W$.

\nref\rcandelas{P. Candelas, Nucl. Phys. {\bf B298} (1988) 458.}%

The expression \eNP\ demonstrates that the numerical value of $\kappa$
depends both on the choice of $\Omega$
and on the coordinate system (in the
language of previous sections $\kappa$ is a section of $\CL^2 \otimes
{\rm Sym}({T^*}^{\otimes d})$).
As we have discussed, a necessary ingredient for the application of
mirror symmetry in this context is to discover the correct
map between $\psi$ and $t$, where the
former is our parameter on the complex structure moduli space
of $M$ and the latter denotes a coordinate on
the K\"ahler moduli space of $\tilde M$.
Choosing a particularly
convenient gauge for $\Omega$
\eqn\egauge{\Omega = \psi\,{{z_1\,dz_2\wedge\dots\wedge dz_{d+1}
+\hbox{cyclic permutations}}\over P} ,}
the techniques of
deformation theory allow us to compute $\kappa_{\psi,\ldots,\psi}$
quite simply \rcandelas\
\eqn\ekapi{\eqalign{
\kappa &= {\psi^2\over H(\psi)}
\ ,\cr
H(\psi) \equiv {\rm det}({\p^2 P\over \p z_i \p z_j}) &=
\left( (d+1)(d+2)\right)^{d+2} (1-\psi^{d+2})\ .\cr
}}
A more natural parameter on the moduli space is $z=\psi^{-(d+2)}$,
and these quantities can equally well be expressed in terms of $z$.
In order to obtain information about rational curves on $W^{(d)}$ we
need to find the correct coordinate $t$ in terms of $\psi$ or $z$.

\nref\rDM{D.R. Morrison, J. Amer. Math. Soc. {\bf 6} (1993) 223.}%

To find this we consider the
periods of the holomorphic $d$-form $\Omega$ along a set of $n$-cycles
locally constant up to homology,
$ \varpi_i =  \int_{\gamma_i} \Omega(z)$.
We restrict the $\gamma_i$'s to lie in the {\it primary horizontal
subspace}\/ of homology, which by definition is the annihilator of
the orthogonal complement of the primary horizontal subspace of
cohomology (introduced in subsection 2.2).
To find the periods in terms of
$\psi$ we will make use of the fact that they satisfy---as discussed
in section II---a set of
differential equations, the Picard--Fuchs equations. For a
one-parameter family this is an
ordinary differential equation with regular singular points at boundary
points of the
moduli space. The monodromy of the locally constant homology cycles
(in the primary subspace) about
these degeneration points is reflected in the monodromy of the solutions.
In particular, the boundary point corresponding to large radius
of $W^{(d)}$ is a singular point of ``maximally unipotent monodromy''
\rDM.
This implies \rDRM\ that a set $\varpi_0, \varpi_1, \dots, \varpi_d$ of local
solutions can be found so that $\varpi_0$ is single-valued,
and each ratio of successive solutions $\varpi_{i+1}/\varpi_i$ has
the form
\eqn\esv{{1\over2\pi i}\,\log z + \hbox{single-valued function}}
near the boundary point.

We then use
\eqn\et{
t = {\varpi_1\over \varpi_0}
\ ,}
to specify the mirror map; the coordinate $q$ in terms of which we
perform power series expansions is then represented as $q = e^{2\pi i t}$.
Note that under transport about the
singular point we have $t\to t+1$. This form of the mirror map
was first advanced by Candelas et al.\ \rCDGP, formulated as described
here in \rDM, and recently explained in \rBCOV.

\nref\rLSW{W. Lerche, D.-J. Smit, and N. Warner,
Nucl. Phys. {\bf B372} (1992) 87.}%
\nref\rsmith{F.C. Smith, Bull. Amer. Math. Soc. {\bf 45} (1939) 629.}%

For the case at hand the  required Picard--Fuchs equations were derived
by Lerche et al.~\rLSW.
The Picard--Fuchs equation is seen to be a generalized
hypergeometric equation (we have set $z = \psi^{-(d+2)}$); the singular
point of interest is $z=0$
\eqn\elsw{
\left[ z\,\prod_{j=1}^{d+1} (z\p_z + {j\over d+1})  -
 (z\p_z)^{d+1}\right] \varpi = 0
\ .}
Using standard techniques (see
\rsmith)
we find the following series expansions for the solutions
\eqn\solns {\eqalign {
\varpi_0 &= \sum_{n\ge 0} \prod_{j=1}^{d+1} \left({j\over d+1} \right) _n
{z^n\over {(n!)}^{d+1}} \cr
\varpi_1 &= \varpi_0 \log (z) + {\p\over\p w}|_{w=0}
\left[\sum_{n\ge 0} \left( \prod_{j=1}^{d+1} {(w+j/d+1)_n\over
(w+1)_n}\right) z^n \right] \ ,\cr
}}
where
\eqn\epoch{
(a)_n \equiv {\Gamma(a+n)\over\Gamma(a)}
}
is the Pochhammer symbol.
Note that \solns\ yields explicit expressions for the power series
coefficients using elementary properties of Gamma functions.

The required series expansion for $\kappa$ is then obtained by inverting
these to express $z$ as a series in $q$ and then inserting \ekapi\ (taking
proper account of the change of coordinates from $\psi$ to $z$).
In table 1 we give the first few terms in these series expansions
of $\kappa_{tt\dots t}$ for
$d$ in the range four to ten.
Notice that, as expected, the series all involve integer coefficients.
However,  it is not immediately clear how to give
geometrical interpretations to these integers.
The key to the explanation is to recall that these $d$-point functions
must factor into sums of products of three-point functions.  As we will
see in the next section, the three-point functions can be directly calculated
for the $\bf B$ model, and the corresponding $\bf A$ model three-point
functions have
an immediate geometrical interpretation which we shall describe.
The power series shown in table 1 will factor into other power series which
explicitly represent these three-point functions.

\subsec{Factorization and Three-Point Functions}

In the previous subsection we computed the $d$-point Yukawa couplings
and found their series expansions. As mentioned there, the objects for
which we have an immediate geometric interpretation are the
three-point functions of the {\bf A} model; this interpretation will
be discussed in detail in section V. In this subsection we will
describe the three-point functions and relate them to the correlators
computed above; section IV is devoted to an algorithm for computing
the three-point functions.

\nref\rDij{R. Dijkgraaf, {\it A geometrical approach to two dimensional
conformal field theory}, Ph.D. thesis, University of Utrecht, pp.~29--34.}%
\nref\rWtop{E. Witten, Nucl. Phys. {\bf B340} (1990) 281.}%

One of the defining properties of a topological field theory is the
factorization property exhibited by its correlation functions. In the
present context this means that all of the correlators can
be written in terms of the nonvanishing two-point and
three-point functions. Underlying this is the fact that the operators
in a topological field theory form an associative, commutative
graded ring on
which the correlation functions determine a trace function \refs{\rDij,\rWtop}.
This is
manifest in the {\bf B} model; the form of \eBcor\ shows that
multiplication in the ring is just the cup product. In the context of
the {\bf A} model this property is less obvious and will lead, after
interpreting the instanton expansion coefficients in terms of rational
curves, to some unsuspected properties of the latter.

The ring structure implies the existence of a topological version of
the operator product expansion, in the form
\eqn\eOPE{
{\cal O}_{\alpha}^{(i)}{\cal O}_{\beta}^{(j)} = {C^{(i,j)}}_{\alpha
\beta}^{\rho} {\cal O}_{\rho}^{(i + j)}
\ .}
Our notation here is that a superscript
$(j)$ indicates that the corresponding operator is in
$H^j(M,\Lambda^j T)$ (for a {\bf B} model computation; the grading
property is universal but not always as obvious)
and the subscripts are labels. Using \eOPE\ it is possible to express
a correlator in terms of correlators with fewer fields. In turn, the
expansion coefficients $C^{(i)}$ themselves may be expressed in terms
of the two-point and three-point correlators.

\nref\rttstar{S. Cecotti and C. Vafa, Nucl. Phys. {\bf B367} (1991) 359.}

This comes about as follows. The two-point
function determines a nondegenerate metric on $H_B$ (of \eBHS )
\eqn\eeeta{
\eta_{\alpha\beta} = \langle{\cal O}^{(i)}_\alpha{\cal O}^{(j)}_\beta\rangle
\ .}
The properties of \eBcor\
guarantee that for arguments $\cal O$ of pure type, $\eta_{\alpha\beta}$ is
nonzero only between complimentary types $i+j=d$.\foot{The metric $\eta$
differs from the metric $G$ discussed in section II, even when restricted
to the subspace of $H_B$ corresponding to marginal deformations; the
relation between these two was the subject of \rttstar.}
The metric $\eta$ depends holomorphically on the parameters and is flat.
In fact, it is possible to choose a basis which varies so that
$\eta_{\alpha\beta}$ is constant. For a one-parameter family, we can restrict
attention to the primary horizontal subspace which is one-dimensional
in each graded piece. We can then certainly choose our normalizations so that
$\eta^{(i,j)}  = c\, \delta_{i+j,d}$ where $c$ is the degree of the
variety.\foot{If we try to suppress this degree $c$ by a change of
basis, then for $d$ even,
in the middle cohomology group $H^{d/2,d/2}$
we would have to leave the realm
of integral cohomology and allow a square root as a coefficient.  For this
reason, we stick with this almost-standard normalization.}

In a similar manner, the three-point functions determine maps
\eqn\eYukmap{
Y_i^j : H^i(M, \Lambda^i T) \times H^j(M, \Lambda^j T)
 \times H^{d - i - j}(M, \Lambda^{d - i - j} T) \rightarrow \IC
}
given in terms of some basis for $H_B$ by%
\foot{For the special case of one-parameter families
in which we focus only on a single element in each $H^p(M, \Lambda^p T)$
we use the same symbol for the map and its image in $\IC$ (for specially
chosen normalization of the arguments).}
\eqn\ealpha{
Y^j_i=\langle {\cal O}^{(i)}{\cal O}^{(j)}{\cal O}^{(d-i-j)}\rangle
=C^{(i,j)}\eta^{(i+j,d-i-j)}
=c\,C^{(i,j)}
\ .}
Because $\eta$ is invertible, we can use this to express $C^{(i,j)}$
in terms of $Y$. This is the sense in which all correlators are
determined by the two-point and three-point functions.
There is an obvious symmetry $Y_i^j = Y_j^i = Y_i^{d-i-j}$ among these
functions. The associativity of the ring of local operators leads to
some less obvious relations which we now discuss.

We now turn to the final goal of this section:
to show that a complete set of three-point functions is provided by those
which involve at least one element in $H^1(M,T)$, i.e.\ the $Y_i^1$.
The essential idea here is that a four-point function can be factored
into (sums of) products of pairs of three-point functions in
up to three distinct ways by using the associativity of the operator
product expansion. To illustrate this point, consider, for example, a four-%
point function $\langle {\cal O}_{\alpha}^{(1)}{\cal O}_{\beta}^{(1)}{\cal
O}_{\gamma}^{(2)}{\cal O}_{\delta}^{(2)}\rangle $ on a Calabi--Yau sixfold.
By factoring this four-%
point function in the two distinct possible ways we have
\eqn\eFactor{ {C^{(1,1)}}_{\alpha \beta}^{\rho} {C^{(2,2)}}_{\delta
\gamma}^{\sigma}
{\cal O}_{\rho}^{(2)}{\cal O}_{\sigma}^{(4)} =
{C^{(1,2)}}_{\alpha \delta}^{\rho} {C^{(1,2)}}_{\beta \gamma}^{\sigma}
{\cal O}_{\rho}^{(3)}{\cal O}_{\sigma}^{(3)}
\ .}
Thus we have
\eqn\eEquality{{C^{(1,1)}}_{\alpha \beta}^{\rho} {C^{(2,2)}}_{\delta
\gamma}^{\sigma} \eta_{\rho \sigma}^{(2,4)} =
{C^{(1,2)}}_{\alpha \delta}^{\rho} {C^{(1,2)}}_{\beta \gamma}^{\sigma}
\eta_{\rho \sigma}^{(3,3)} .}
We see from this equality that if we know the metric $\eta$ and the
operator product coefficients $C^{(1,1)}$ and $C^{(1,2)}$, then
associativity gives us a set of linear equations for the coefficients
$C^{(2,2)}$. In the normalization discussed above we can make this
more explicit and find
\eqn\eRel{C^{(1,1)}C^{(2,2)} = (C^{(1,2)})^2 \ .}

Using \ealpha , \eRel\ gives a relation
\eqn\eRell{ Y^2_2 =(Y^1_2)^2/Y^1_1 .}
This same reasoning is readily used to show that for arbitrary $d$
(in the primary horizontal subspace)
 we
have
\eqn\eGeneral{ Y^j_i = \prod_{k = 0}^{j-1} Y^1_{i+k} / \prod_{k=1}^{j-1}
Y^1_k.}
We thus see that all Yukawa couplings $Y^j_i$ are determined in terms of those
which contain at least one member of $H^1(M,T)$.

As discussed above, the $Y^j_i$ are interpretable as three-point functions
on the mirror $\tilde M$ involving elements of $H^i(\tilde M, \Lambda^i T^*)$,
$H^j(\tilde M, \Lambda^j T^*)$, and $H^{d-i-j}(\tilde M, \Lambda^{d-i-j} T^*)$.
These three-point functions have instanton expansions whose coefficients
depend on the rational curves on $\tilde M$. The identities in \eRell\
(and their straightforward generalizations to higher dimensional moduli
spaces) thus provide various relations among the numbers associated to
rational curves. These relations provide a sensitive consistency check
on our methods (as we shall see).

\newsec{The Mirror Map and Three-Point Functions}

As discussed earlier, the arguments of \rGP\ establish an
abstract isomorphism between the moduli spaces of complex
structures on $M$ and K\"ahler structures on $\tilde M$ and between
the associated Hilbert spaces which preserves the correlation functions.%
\foot{The argument in \rGP\ establishes this up to possible global
considerations.} A full understanding of mirror symmetry, and certainly
its application to computing properties of rational curves, requires
knowing
the explicit form of these isomorphisms. As mentioned in the previous
section, an ansatz for the so-called ``mirror map'' between the moduli
spaces was proposed (and verified in an example) in \rCDGP ; this has
since been checked in many other examples and has recently been
explained in \rBCOV. This map provides naturally a part of the
required isomorphism between the Hilbert spaces $H^p(M,\wedge^p T)$
and $H^p(\tilde M,\wedge^p T^*)$, since the tangent directions to moduli
space are related to the subspace of marginal operators (recall this
is simply the subspace at the first nonzero grading). This
isomorphism was used in the previous section to relate correlators of
these operators in the two models, and the fact that the series of
table 1 yield integer coefficients is a signal that we have performed
the mapping correctly. As we have seen, however, these correlators are
in some sense secondary objects derived from the more fundamental
three-point couplings; it is to these fundamental objects that a
geometrical interpretation (in the {\bf A} model) may be given. These
however necessarily involve non-marginal operators, so we will need to
extend the mirror map to a complete isomorphism between the Hilbert
spaces. In the special case of
 Calabi--Yau threefolds, mapping the space of marginal operators
in fact suffices to extract
all of the required information. As mentioned in section II, complex
conjugation generates from a basis of these a basis for the entire
space; performing this operation in both spaces leads to two bases
related by mirror symmetry.
In  other cases, however, constructing
bases (as sections over moduli space) which are mapped to each other
by mirror symmetry requires more structure. We will supply this
structure and  give a systematic method for finding such
bases and hence exploiting mirror symmetry. We will focus our attention
on one-parameter families of Calabi--Yau $d$-folds in projective
space although the extension to weighted projective spaces and
higher dimensional moduli spaces should be relatively straightforward.

As a brief summary for the rest of this section, we note that our
approach is, roughly, as follows. The mirror map, as discussed in
section III, determines a coordinate (the ratio of periods) on
the moduli space of complex structures on $M$ related by
mirror symmetry to the natural coordinate on the space of K\"ahler
structures on $\tilde M$.
Our goal is to use this information, which essentially gives us
mirror symmetric bases of $H^1(M, T^*)$ and $H^1(\tilde M, T)$
to construct mirror symmetric
bases of
$H^p(M,\wedge^p T)$ and $H^p(\tilde M,\wedge^p T^*)$.
In essence, we construct such bases by beginning with elements in the
$H^1$ cohomology groups and generating the primary subspace by
successive operator products of these.
On $\tilde M$ we will relate this to
an integral basis of $H^{p,p}(\tilde M, \BZ)$. On $M$,
we find a systematic approach using the Gauss--Manin connection.

\subsec{The Gauss--Manin Connection and the Choice of Basis}

The Gauss--Manin connection $\nabla$ was introduced in section II as
the flat holomorphic connection compatible with the metric on $\cal H$.
This connection can also be defined by the following important property.
As we move around in the parameter space of complex structures of $M$,
the decomposition $H^d(M,\IC) = \oplus_p H^{p,d - p}(M,\IC)$ varies since the
meaning of a $(p,q)$ form depends upon the complex structure.
We can, however, also consider a topological basis of $H^d(M,\IC)$ (for
example, the  duals of topological homology cycles in
$H_d(M,\IC)$) which does not vary with the complex structure. The Gauss--Manin
connection measures the variance of the former basis with respect to the
latter.
To see this explicitly, let $\gamma_1$, \dots, $\gamma_k$ be
a topological basis of $H_d(M,\IC)$ and consider $\alpha(z)$ to be a
holomorphically varying element in ${\cal F}^p$. Then, we can write
\eqn\eExpand{ \alpha(z) = \sum_\mu ( \int_{\gamma_\mu} \alpha(z) ) \gamma_\mu^*
}
with $\{\gamma_\mu^* \}$ being the  dual basis of $\{\gamma_\mu\}$ in
$H^d(M,\IC)$.
We define the action of $\nabla$ to be
\eqn\eGM{ \nabla \alpha =  \sum_{i,\mu} ( \int_{\gamma_\mu}
\del_{z_i} \alpha(z) ) \gamma_\mu^* dz^i .}
In other words, the Gauss--Manin connection is defined by demanding that the
topological sections $\gamma_\mu$ are flat sections. Then, covariant
differentiation turns into ordinary differentiation with respect to the
parameters of the complex structure moduli space.
We will momentarily see that the Gauss--Manin connection plays a crucial role
in finding and implementing the extended mirror map.

We now, once again, specialize our discussion to the case $h^{d-1,1}_M =
h^{1,1}_{\tilde M} = 1$ and to the primary horizontal and vertical
subspaces of $H^{j, d - j}(M,\IC)$
and $H^{j,j}(\tilde M ,\IC)$ generated by these one-dimensional
spaces. Our goals
are to

1)  find a map from the moduli space ${\cal M}^{c.s.}_M$
 of complex structures
on $M$ parameterized by the complex coordinate $z$ to the ``K\"ahler''
moduli space
${\cal M}^{K}_{\tilde M}$  parameterized by the complex coordinate $t$
and to

2)  find the explicit isomorphism between
$\oplus_p H^{p , p}(\tilde M,\IC)$
and $\oplus_p H^{p , d - p}(M,\IC)$

\noindent
such that the {\bf A} model Yukawa couplings
$\langle \tilde{\cal O}^{(i)}\tilde{\cal O}^{(j)}\tilde{\cal O}^{(k)}\rangle$
as functions of $t$ are equal to the {\bf B} model Yukawa couplings
$\langle {\cal O}^{(i)}{\cal O}^{(j)}{\cal O}^{(k)}\rangle$
as functions of $z$ (for corresponding basis elements) once we express $t$
in terms of $z$ using the mirror map.

To this end, we first note that there is an especially convenient basis
for the primary vertical subspace of $\oplus_p H^{p , p}(\tilde M,\IC)$.
It can be described as $e_0,e_1,\dots,e_d$, where each $e_p$ is the
integral generator of $H^{p , p}(\tilde M,\IC)$ which is the Poincar\'e
dual of a submanifold of complex codimension $p$.
(We in fact take $e_p$ of the form $e_1\cup{\cdots}\cup e_1$ (with $p$ terms).)
As discussed earlier, it is this basis which gives rise
to the simplest geometrical interpretation of three-point
correlation functions. Goal (2) will be
achieved if we can find the mirror image  of this basis in
$H^d(M_z , \IC)$.
Moreover, since the K\"ahler moduli space of $\tilde M$ is locally
isomorphic to $H^1(T^*_{\tilde M})=H^{1,1}(\tilde M)$, the generator
$e_1$ of $H^{1,1}(\tilde M)$ determines a natural coordinate $t$
on the K\"ahler moduli space.  (The K\"ahler form will be written as
$t\, e_1$.)
 So we can actually achieve both
 goals $(1)$ and $(2)$ by
finding the appropriate analogous basis in $H^d(M_z , \IC)$,
since the analog of $e_1$ can be used to specify a coordinate.

To motivate our solution to this question, let's look more closely at the
the primary vertical sub-basis $e_0,e_1,\dots,e_{d}$ of
$\oplus_p H^{p , p}(\tilde M,\BZ)$. We
have, in this basis,
\eqn\eETA{\eta^{(i,j)} = \langle e_i, e_j\rangle  = c\, \delta_{i+j,d} ,}
where $c$ is a specific constant, the {\it degree\/} of
$\tilde M$, calculated by integrating $e_1\cup{\cdots}\cup e_1$ (with $d$
terms)
over $\tilde{M}$.\foot{In odd dimension we could change basis to get rid of
this constant, but in even dimension doing so would introduce the
square root of the degree as a coefficient, which could provide a good
basis for {\it real\/} cohomology but not for {\it integral\/} cohomology.}
Also note that we clearly have $e_1 e_d = 0$. Our basis, therefore,
satisfies the following three features:

1) $ e_1 e_{j-1} =  {{ c^{-1} A^1_{j-1}(t) }} \ e_j$

2) $\eta^{(i,j)} = \langle e_i, e_j\rangle  = c\,\delta_{i+j,d}$

3) $e_1 e_d = 0$.

\noindent
where we have used $A^1_{j-1}(t)$ to denote the {\bf A} model Yukawa
coupling $\langle e_1e_{j-1}e_j\rangle$ as a function of $t$.\foot{These
functions coincide with the function $Y^1_{j-1}$ of subsection 3.2; we
introduce the notation $A^1_{j-1}$ here and $B^1_{j-1}$ below in order
to emphasize when these functions are being calculated on the {\bf A}
model of $\tilde M$, and when on the {\bf B} model of $M$.}

Of course, property (1) follows from (2), but writing it in this
manner will be useful shortly.
In particular,
we interpret (1) as follows:  the operator product of $e_1$ and $e_{j-1}$
is a functional multiple of $e_j$, with the multiplier depending on
the parameter in the K\"ahler moduli space.
We note that on the $\bf A$ model side
these properties do not uniquely single out
a basis; rather, they are properties characteristic of a set of bases,
amongst which is the basis of integral generators.

\nref\rGRIFF{P.A. Griffiths, Ann. of Math. {\bf 90} (1969) 460.}%

We now mimic these properties on the $\bf B$ model side; we will see
that a slightly stronger version of these properties,
when combined with an analysis of the monodromy, does determine an
essentially unique basis.

We will formulate our basis for the $\bf B$ model bundle in such a
way that both the basis and the correlation functions manifestly have
a holomorphic dependence on moduli.  As has been recognized since
the work of Griffiths \rGRIFF, there is an inherent conflict between
choosing bases of pure $(p,q)$ type, and choosing bases which vary
holomorphically with moduli.  (This is why we introduced the bundles
${\cal F}^p$ rather than working directly with $H^{p,q}(M)$
in our discussion of the Gauss--Manin connection, since
the ${\cal F}^p$'s are the holomorphically varying objects.)  Although
the first choice might appear at first sight to be better adapted
to a study of mirror symmetry
(since we usually work on the $\bf A$ side
with bases of pure type), the holomorphic dependence of $\bf B$
model correlation functions is difficult to see if calculations are
made in a non-holomorphic gauge.  So we adopt the second strategy, and
abandon pure type in favor of holomorphically varying bases.
At the end of the analysis, we can obtain a basis of pure type by
simply projecting to the appropriate $(p,q)$ pieces.

At a single point in the moduli space, the $\bf B$ model three-point functions
\eqn\eag{B^1_{j-1}:H^1(M,T) \times H^{j-1,d-j+1}\times  H^{d-j,j}\to\IC}
have a natural description in algebraic geometry coming from ``variation
of Hodge structure'':  they describe what is called
the ``differential of the period map''.
In fact, identifying $H^1(M,T)$ with the tangent space
to the complex
structure moduli space at the point corresponding to $M$,
this three-point function describes the
$(p-1,q+1)$ part of a derivative (with respect to parameters) of a
family of $(p,q)$ forms (taking $p=j-1$ and $q=d-j+1$, say).
The Gauss--Manin connection introduced above
reproduces this derivative information while preserving holomorphic
dependence.  The result of a ``pure type'' differentiation may differ
from the Gauss--Manin answer by some terms of lower type, but all such
terms vanish after
wedging with a $(d-p+1,d-q-1)$ form and integrating (the prescription for
calculating the three-point function).

\nref\rDG{J. Distler and B. Greene, Nucl. Phys. {\bf B309} (1988) 295.}%

Let us fix a holomorphic vector field $\xi$ in the moduli space in such
a way that the directional Gauss--Manin derivative  $\nabla_\xi\Omega$
produces a chosen initial
basis vector $\alpha_1(z)$ in $H^{d-1,1}$.
Then one can prove that the following two operations are identical:

i) taking the directional Gauss--Manin derivative $\nabla_\xi$
and projecting onto the $(p,q)$ term in the result, for largest q

ii) taking the operator product with the $(\rm chiral, \rm chiral)$
field
$\alpha_1$ of charge $(1,1)$ corresponding to $\xi$

\noindent
This fact can be established by the methods of \rCd\ along with
the nonrenormalization theorem of \rDG\ which establishes the equality
of operator products amongst $(\rm chiral, \rm chiral)$ fields and
standard mathematical wedge products.

The operation (i), however, does not respect holomorphicity (as a function
of the moduli space coordinate), as we have noted. Holomorphicity
requires that we do not project the result onto the term of highest
antiholomorphic degree. On the other hand, agreement with the conformal
field theory operator product demands that we do. It appears that
essentially all correlation functions, though, are insensitive to
these additional lower order terms which are responsible for
holomorphicity. Hence, by including
these terms we gain the benefit of holomorphically varying elements
(as we do on the $\bf A$ side)
without altering the values of correlation functions.
Thus, the central assumption  of our analysis is that we
construct the basis on the $\bf B$ side by imposing
the same three conditions as on the $\bf A$ side
{\it with the replacement of the operator product by the action
of the (unprojected) Gauss--Manin connection}.\foot{A basis consisting
of forms of pure type can then be obtained from the basis we construct
by simply projecting each basis element to the appropriate $(p,q)$ piece.}

That is, we build our basis of the horizontal subspace of
 $\oplus_p H^{p , d -p}(M,\IC)$ by
beginning with
$\alpha_0=\Omega$, and then seeking $\{\alpha_j\}$ such that

$1'$) $ \nabla_{\alpha_1} \alpha_{j-1}  =  c^{-1} B_{j-1}^1 \ \alpha_j ,$

$2'$) $\langle \alpha_i,\alpha_j\rangle  = c\, \delta_{i+j,d}$, and

$3'$) $\nabla_{\alpha_1}\alpha_d = 0$.

\noindent
where $c$ is an appropriately chosen constant (which will correspond to the
degree of the mirror variety).
Note that $B_0^1$ is constant, and equal to the degree $c$, so that
$\nabla_{\alpha_1}\alpha_0=\alpha_1$ provides the link between the
directional derivative $\nabla_{\alpha_1}$ and the form $\alpha_1$.
We also note that
were we to use the projected Gauss--Manin derivative, condition
$3'$ would be trivial (as is its counterpart condition $3$).
However, because we use the (unprojected) Gauss--Manin derivative,
which does not yield results of pure type, our $\bf B$ model
conditions are
 somewhat more
stringent than their $\bf A$ model counterparts. This will manifest
itself in the solutions to these conditions being essentially
unique, unlike the case on the $\bf A$ side.
(At first sight it might appear asymmetric to begin with $\alpha_0 =
\Omega$ since on the $\bf A$ side we begin with $\e_0 = 1$. This
is just an artifact of our working  in ${\cal H}$ rather than in
${\cal H} \otimes {\cal L}^{-1}$, as we have discussed earlier,
in which $\Omega$ can be thought of as $\boldone = \Omega \otimes
\Omega^{-1}$. In fact,
we will shortly find it convenient to essentially divide by $\Omega$
in a similar manner.)

To find a basis meeting these conditions it proves convenient to
introduce a set of topological homology cycles $\gamma_0, \gamma_1, \dots,
\gamma_d$ spanning the primary horizontal subspace,
such that the cup product pairing on the dual cohomology cycles $\gamma_\mu^*$
satisfies
\eqn\ePairing{
(\gamma_\mu^*,\gamma_\nu^*)=\cases{ 0 & \hbox{\quad if $\mu+\nu\gt d$}\cr
c & \hbox{\quad if $\mu+\nu=d$}}
}
(there is no constraint on the values when $\mu+\nu\lt d$).
We can then express our basis $\alpha_i$ in terms
of the $\gamma_\mu$ by writing the ``period matrix''
\eqn\eXX{
P=(P_{i\mu}) = (\int_{\gamma_\mu}\alpha_i).
}
(Indices on matrix elements run from $0$ through $d$.)
We claim that we can achieve constraints ($2'$) and ($3'$) by performing row
operations to put this matrix in upper triangular form with the diagonal
entries being all one, that is, achieving the conditions
\eqn\eUT{
\int_{\gamma_\mu}\alpha_i=\cases{ 0 & \hbox{\quad if $i\gt\mu$}\cr
1 & \hbox{\quad if $i=\mu$}}.
}
The row operations we allow include adding one row to a later row, and
multiplying a row by an arbitrary holomorphic function of $z$.  (It
is clearly necessary to allow this last step, if we are to achieve
$\int_{\gamma_i}\alpha_i=1$.)
These row operations effectively alter the basis $\{\alpha_j\}$, but they
do preserve the property $\alpha_j\in{\cal F}^{d-j}$.  Note that the
use of holomorphic bundles ${\cal F}^{d-j}$ was crucial here, since
we must allow arbitrary holomorphic functions as multipliers.

To see that ($1'$) holds for this new basis is straightforward.  Writing
\eqn\ealphaexpand{\alpha_{j-1}=\gamma_{j-1}^*+\sum_{\ell=j}^d\left(
\int_{\gamma_\ell}\alpha_{j-1}\right)\gamma_\ell^*}
we find
\eqn\egradalpha{\nabla_{\alpha_1}\alpha_{j-1}=
\sum_{\ell=j}^d\,{d\over d\alpha_1}\left(\int_{\gamma_\ell}\alpha_{j-1}\right)
\gamma_\ell^*.}
This is an element of ${\cal F}^{d-j}$, and so must be a linear combination
of $\alpha_0$, \dots, $\alpha_j$.  It follows from \eUT\ that the
coefficient of $\alpha_0$ in the linear combination should agree with
the coefficient of $\gamma_0^*$ in \egradalpha, but this is zero.
That being the case, the coefficient of $\alpha_1$ in the linear combination
should agree with the coefficient of $\gamma_1^*$ in \egradalpha, but
this too is zero.  Continuing to argue
in this way we find that $\nabla_{\alpha_1}\alpha_{j-1}$
must simply be a multiple $f_{j-1}(z)\cdot\alpha_j$.  In fact, the multiplier
is easily seen to be
\eqn\emultiplier{f_{j-1}(z)={d\over d\alpha_1}\int_{\gamma_j}\alpha_{j-1}.}

Condition ($3'$) holds as it translates into
the covariant derivative of the last row of the matrix vanishing---this
is clearly true as the last row of the matrix is constant. To check
condition
($2'$), first note that because
we have preserved the condition
$\alpha_j\in {\cal F}^{d-j}$,
by considering types in the wedge product we find
\eqn\evanish{
\langle \alpha_i,\alpha_j\rangle =0\hbox{\quad if $i+j\lt d$.}
}
Thus, we may assume $i+j\ge d$.  We then calculate
\eqn\eCalc{
 \langle \alpha_i,\alpha_j\rangle =
\sum_{\mu,\nu}\left(\int_{\gamma_\mu} \alpha_i\right)
\left(\int_{\gamma_\nu} \alpha_j\right)\,(\gamma_\mu^*,\gamma_\nu^*).
}
For any term in this last sum which is non-zero, we must have
\eqn\eIneqs{
d\le i+j\le\mu+\nu\le d
}
(using \eUT\ and \ePairing).  Thus, all inequalities are equalities,
and we find
\eqn\eCalcc{
 \langle \alpha_i,\alpha_j\rangle =
\sum_{\mu,\nu}\delta_{\mu i}\delta_{\nu j}(c\,\delta_{\mu+\nu,d})
=c\,\delta_{i+j,d},
}
as required.

(As one final check, we can evaluate the three-point function
\eqn\etpf{B_{j-1}^1=\langle \alpha_1\alpha_{j-1}\alpha_{d-j}\rangle
=f_{j-1}(z)\langle \alpha_j,\alpha_{d-j}\rangle =c\cdot f_{j-1}(z),}
so $f_{j-1}(z)=c^{-1}B_{j-1}^1$ as asserted in ($1'$).)

Notice that in performing row operations to make  the matrix $P$
upper triangular with $1$'s on the diagonal,
the only manipulation which affected the top row divided it
by $\int_{\gamma_0} \Omega(z)$ thereby making the $(0,0)$ entry
in the new matrix
equal to $1$ and the $(0,1)$ entry
equal to
$\int_{\gamma_1} \Omega(z)  \over \int_{\gamma_0}
\Omega(z).$ As the derivative of the top row with respect to
$t$ is the second row, and since the $(1,1)$ entry is $1$, we
directly see that
in our new basis, $\nabla_{\alpha_1} = \partial_t$ with
\eqn\eCOORD{t = {  \int_{\gamma_1} \Omega(z)  \over \int_{\gamma_0}
\Omega(z)}.}
This is precisely the same coordinate {\it ansatz\/} used in \rCDGP\
and established in \rBCOV\ as being mirror
to the {\it integral\/} generator of $H^2(\tilde M)$; we see here
that this form of the mirror map {\it emerges\/} from
our three conditions.  Although our conditions
on the $\bf A$ side do not uniquely specify a basis, as we discuss below,
our slightly stronger conditions on the $\bf B$ side, combined with
monodromy properties, make the basis essentially unique. Since our
procedure on the $\bf B$ side has picked out the first element of
this basis
to be the known mirror of an integral generator, we expect that
the same is true for the other elements of the $\bf B$-basis, as
desired.

Having now satisfied the characteristics of the $\bf A$ model
basis for the primary vertical subspace of $\oplus_p H^{p , p}(\tilde M,\BZ)$
with the $\bf B$ model basis of the primary horizontal subspace of
$\oplus_p H^{p , d -p}(M,\IC)$ (under our central assumption
discussed above), we now
must ask ourselves about the uniqueness of this procedure.
The first point to make about uniqueness is this:  any basis which
satisfies our conditions ($1'$), ($2'$) and ($3'$) must also satisfy
\eUT\ for {\it some\/} choice of homology cycles $\gamma_\mu$.
This can be seen as follows.  We start with an arbitrary basis
$\gamma_0$, \dots, $\gamma_d$ of the primary horizontal subspace
and form the period matrix \eXX\ with
respect to that basis.  We then perform column operations on this matrix to put
it into upper triangular form with $1$'s on the diagonal, but this time
we restrict ourselves to using {\it constants\/} as multipliers  for
the columns.  (This has the effect of changing the basis $\gamma_\mu$,
using linear combinations with constant (complex) coefficients.
Under such a change, the $\gamma$'s
 will remain a basis of the primary subspace of $H^d(M,\IC)$.)
We are aiming for the condition \eUT, but since we have restricted our
allowed multipliers it would seem problematic to achieve
$\int_{\gamma_i}\alpha_i=1$.

However, conditions ($1'$) and ($3'$) come to our rescue.  First,
($3'$) implies that the bottom row of $P$ is {\it constant}.
Therefore, by suitable constant-coefficient column operations we
can put the bottom row in the form
\eqn\ebotrow{\pmatrix{0& 0& \cdots& 0& 1}.}
It then follows from ($1'$) with $j=d$  that every entry but the last one
in the penultimate row is constant.  Again applying
constant-coefficient column operations (which do not involve the last
column) we can achieve for the bottom two rows:
\eqn\ebottwo{\pmatrix{0&0&\cdots&0&1&\star\cr0&0&\cdots&0&0&1}}
where $\star$ is an unknown quantity.  Continuing in this way
row by row produces \eUT.

Although this argument eliminates the apparent arbitrariness of using
the condition \eUT\ to achieve ($1'$), ($2'$) and ($3'$), it still
leaves us with a procedure that is not unique---the starting set of cycles
$\{\gamma_\mu\}$ used to produce the basis $\{\alpha_j\}$ is not unique.
We can, however, make this choice essentially unique by going to a boundary
point in the moduli space. As discussed in \refs{\rCDGP, \rDM}, the cycles
$\gamma_\mu$ have nontrivial monodromy about boundary points in the moduli
space. We also know, from the $\bf A$ model, that at a large radius boundary
point we have the identification of $t$ and $t + 1$. Thus, consistency of the
mirror map will follow if
the monodromy of the $\gamma_\mu$ is ensured to yield
$ {  \int_{\gamma_1} \Omega(z)  \over \int_{\gamma_0} \Omega(z)} \rightarrow
{  \int_{\gamma_1} \Omega(z)  \over \int_{\gamma_0} \Omega(z)}  + 1 $.
This is sufficient to almost uniquely fix the cycles and hence our procedure
for generating the mirror map, as we shall now show.

On the $\bf A$ model side, the physics is the same at $t+1$ as it is at
$t$, and the quantity $q=e^{2\pi i\,t}$ serves as the natural parameter
(near the boundary) on the true moduli space of physical theories.
On the $\bf B$ model side, our monodromy property effectively means
\eqn\emonprop{{\int_{\gamma_1}\Omega(z)\over\int_{\gamma_0}\Omega(z)}
={1\over2\pi i}\log z+f(z)}
for some single-valued function $f(z)$:  the ``$t$'' type parameter is
$\int_{\gamma_1}\Omega(z)/\int_{\gamma_0}\Omega(z)$ while the ``$q$''
type parameter is the exponential of this.
Our directional derivative $\nabla_{\alpha_1}$ (which is being identified with
the mirror of the operator product with $e_1$) behaves like
${d\over dt}=q\,{d\over dq}$ near the large complex structure limit.
In particular, since the three-point functions
\eqn\etpfq{\int q{d\over dq}(\alpha_{j-1}\wedge\alpha_{d-j})
=c^{-1}\cdot f_{j-1}(q)=c^{-1}\cdot q{d\over dq}\int_{\gamma_j}\alpha_{j-1}}
have expansions of the form
\eqn\etpfexp{a_{j-1,0}+a_{j-1,1}\,q+a_{j-1,2}\,q^2+\cdots}
(consisting of a topological term plus quantum corrections), we see that
whenever $a_{j-1,0}\ne0$, the quantity ${d\over dq}\int_{\gamma_j}\alpha_{j-1}$
must have a pole at $q=0$:  the leading term in its Laurent expansion will
be $c\cdot a_{j-1,0}\,q^{-1}$.  Thus the period $\int_{\gamma_j}\alpha_{j-1}$
will have the form
\eqn\eperiodexp{\int_{\gamma_j}\alpha_{j-1}=
c\cdot a_{j-1,0}\,\log q + \hbox{single-valued function}.}

Now we know that the topological terms in these three-point functions cannot
vanish, since they give the degree of the variety, which is nonzero.  Thus,
every entry in the first superdiagonal of the period matrix has a $\log q$
type monodromy.  This is a very strong property, called {\it maximally
unipotent\/} in \rDM.

In the presence of maximally unipotent monodromy, we need a basis
$\gamma_0$, \dots, $\gamma_d$ such that the monodromy action takes the form
\eqn\emonaction{
\gamma_\mu\mapsto\gamma_\mu + \sum_{\nu\lt\mu} m_{\mu\nu}\,\gamma_\nu}
for some constants $m_{\mu\nu}$.  Moreover, our basis should satisfy
\ePairing; these two properties together fix
the $\gamma_\mu$'s up to scalar multiples.

Notice that although our procedure for generating the mirror map and
the appropriate basis in the $\bf B$ model required that we start
with $\alpha_0$ equal to some holomorphic three form $\Omega(z)$,
in reality the particular  initial choice of $\Omega$ is irrelevant,
as we quickly indicated earlier.
Directly we see this as our three conditions lead us to rescale
$\alpha_0$ by $1/\int_{\gamma_0} \Omega(z)$. Alternatively, we could
rephrase all of our analysis along the lines of section II in which
we work in the context of ${\cal H}\otimes{\cal L}^{-1}$ rather than
$\cal H $. As discussed in that
section, the analysis can be phrased as starting with the
canonical section $\boldone$ of ${\cal O}\subset{\cal H}\otimes{\cal L}^{-1}$,
thus ensuring that the results do not
depend on any initial choice of $\Omega$. This approach is closer, in fact,
to our $\bf A$ model description because in that setting we choose
$e_0 = \boldone$ and, furthermore,
because the fibers of ${\cal H}\otimes{\cal L}^{-1}$ are
canonically isomorphic to $H^p(M,\Lambda^p T)$. The latter, as we
have discussed, is the precise geometrical description of the
$(c,c)$ ring, just as $H^p(M,\Lambda^p T^*)$ is that for the
$(a,c)$ ring.

It is worthwhile reemphasizing  that the basis
elements $\alpha_i$ which
we have derived here are generally of mixed type. This is due to our
implicit requirement that the basis be holomorphically varying over
moduli space. It is straightforward to see that it is only the
$(p, d-p)$ part with largest $p$ contained in each $\alpha_i$
that contributes to correlation functions. Thus, if we are willing
to sacrifice holomorphic variation we can eliminate the lower order
pieces.
Such a {\bf B} model basis would more closely match the {\bf A} model
analysis.  Alternatively, we could modify the {\bf A} model basis
to behave more like the holomorphically varying {\bf B} model basis.

There is an added bonus to our procedure beyond
naturally generating the
mirror map and mirror bases.
 The fundamental three-point functions $Y^1_j$
(and their associated instanton expansions) can be directly
extracted from the matrix \eXX. This is easily seen by noting that
the three-point function $Y^1_j$ can be expressed as
\eqn\eYukalt{ Y^1_j(\alpha_1, \alpha_j, \alpha_{d-j-1} )
 = \int_{M_z} \alpha_{d-j-1} \wedge\nabla_{\alpha_1} \alpha_j .}
Substituting in the basis which puts $P$ into upper triangular form, we
directly
calculate that
\eqn\eYUkderiv { Y^1_j = c\cdot \del_t (P_{j,j+1})  .}
Let us reemphasize that these three-point functions, although calculated on
$M_z$, are now to be thought of as three-point functions on
$\tilde M_{t(z)}$. Since we have carefully extracted the mirror map
and identified the bases of cohomology on both sides, \eYUkderiv\ can
directly be interpreted as an instanton sum as in \eXX.

We apply this formalism to specific examples in the next subsection.

\subsec{Holomorphic Picard--Fuchs Equation and Three-Point Functions}

We now employ the discussion of the last subsection to calculate
all of the three-point functions and their associated instanton sums
for the independent set of Yukawa couplings $Y_j^1$ for the
mirror manifolds built on the $M_{\psi}^{(d)}$ introduced in section III.

In practice, we carry out the procedure of the last subsection as follows.
We have $M_{\psi}^{(d)}$
 described by the equation \eFamily, where we are using
the coordinate $z=\psi^{-(d+2)}$ on the moduli space.  One can directly
check from the Picard--Fuchs equation \elsw\ that the only point in
the moduli space with maximally unipotent monodromy is the point $z=0$.
We adapt the methods of \rDRM\ to do our calculation at that point.

We take as our initial basis of the horizontal subspace of
 ${\cal F}^0$ the $d$-forms
\eqn\edforms{\alpha_0=\Omega,\ \alpha_1=z\del_z\,\Omega,\ \cdots,\
\alpha_d=(z\del_z)^d\,\Omega.}
The differentiation operator $z\del_z$ acts on this basis via a matrix
of the form
\eqn\eaction{
 A(z) = \pmatrix{
    0    &    1    &        &        &            \cr
         &    0    &   1    &        &            \cr
         &         & \ddots & \ddots &            \cr
         &         &        &   0    &    1       \cr
 B_0(z)  & B_1(z)  & \dots  & \dots  & B_{d}(z)
},}
where the $B_j(z)$ are determined from the Picard--Fuchs equation \elsw\
as follows \rDRM.  Write the Picard--Fuchs operator in the form
\eqn\epfop{
 z\,\prod_{j=1}^{d+1} (z\p_z + {j\over d+1})  -
 (z\p_z)^{d+1}
=(z-1)(z\p_z)^{d+1} + z\,\sum_{j=0}^d c_j(z\p_z)^j
}
and divide by $z-1$ to produce the operator
\eqn\eop{(z\p_z)^{d+1}-\sum_{j=0}^d c_j{z\over1-z}(z\p_z)^j.}
Then the entries in the bottom row of the matrix $A(z)$ are the quantities
$B_j(z)=c_j\,{z\over1-z}$.  Note that $B_j(0)=0$, so that the methods
of \rDRM\ can be directly used to solve the equation.

For any homology cycle $\gamma$ in the primary horizontal subspace, the vector
\eqn\evector{\varpi(z)=\pmatrix{\int_\gamma\Omega\cr
\int_\gamma z\p_z\Omega\cr\vdots\cr\int_\gamma(z\p_z)^d\Omega}}
is a solution to the matrix equation
\eqn\emateqn{z\p_z\varpi(z)=A(z)\,\varpi(z).}
Most of these solutions are multiple-valued; the multiple-valuedness
can be accounted for in advance as follows.  Our desired basis of homology
cycles $\gamma_0$, \dots, $\gamma_d$ will have the property that
$\int_{\gamma_0}\Omega$ is single-valued, and
\eqn\ebasisprop{2\pi i\,{{\int_{\gamma_j}\Omega\over\int_{\gamma_{j-1}}\Omega}}
=\log z + \hbox{single valued function}.}
We take the corresponding vectors $\varpi_j(z)$ which are solutions to
\emateqn\ and arrange them as columns in a matrix $\Phi(z)$.  This
matrix of multiple-valued functions
satisfies the equation $z\p_z\Phi(z)=A(z)\,\Phi(z)$.  In addition,
there is a matrix $S(z)$ with single-valued entries such that
\eqn\ePhi{\Phi(z)=S(z)\cdot z^{A(0)}}
where $z^{A(0)}$ denotes
$e^{(\log z) A(0)}=I+(\log z)A(0)+{1\over2!}(\log z)^2A(0)^2+\cdots$.
The equation satisfied by $S(z)$ is
\eqn\eSeqn{z\p_zS(z)+S(z)\cdot A(0)=A(z)S(z)}
(see \rDRM).

In our case, the matrix $A(0)$ takes a particularly simple form
\eqn\eAsimple{A(0)=\pmatrix{
0 & 1 &  &  &         \cr
  &   0    &       1         &  &   \cr
  &        &         0            & \ddots &                        \cr
  &        &                      & \ddots &           1             \cr
  &        &                      &        &             0
},}
which leads immediately to
\eqn\ezA{z^{A(0)}=\pmatrix{
1 & \log z & {1\over2!}(\log z)^2 & \cdots &     {1\over d!}(\log z)^d    \cr
  &   1    &       \log z         & \cdots & {1\over(d-1)!}(\log z)^{d-1} \cr
  &        &         1            & \ddots &           \vdots             \cr
  &        &                      & \ddots &           \log z             \cr
  &        &                      &        &              1
}.}
Also thanks to the special form of $A(0)$, the equation \eSeqn\ can
be written as
\eqn\eSeqnn{z\p_z\sigma_j(z)+\sigma_{j-1}(z)=A(z)\,\sigma_j(z),}
where $\sigma_0(z)$, \dots, $\sigma_d(z)$ are the columns of $S(z)$
(setting $\sigma_{-1}(z)=0$).  Solutions to equations \eSeqnn\ can then
be found by power series techniques.

The next step is to put the solution matrix $\Phi(z)$ into upper
triangular form with $1$'s on the diagonal
by means of row operations.  Since $z^{A(0)}$
is upper triangular with $1$'s on the diagonal,
it suffices to put $S(z)$ into upper triangular
form with $1$'s on the diagonal.  This is a straightforward manipulation
with power series, and produces a matrix $\tilde{S}(z)$.  We then have
\eqn\eMSz{P=\tilde{S}(z)\,z^{A(0)}}
where $P$ is the period matrix \eXX.

Using \ezA\ and \eMSz, we deduce
that $P_{j,j+1}=\log z + \tilde{S}_{j,j+1}$.
(Our matrix indices still run from $0$ through $d$.)
Thus, the Yukawa coupling is given by \eYUkderiv
\eqn\eYonej{Y_j^1=c\cdot(1+z\p_z\tilde{S}_{j,j+1})\cdot z\p_zt}
where $c$ is the degree.  (The factor of $z\p_zt$ is present to change from
$z\p_z$ gauge to $\p_t$ gauge.)  Since $Y_0^1=c$, we can solve for the
change of gauge
\eqn\ecog{z\p_zt={1\over1+z\p_z\tilde{S}_{0,1}}}
and find that
\eqn\eYonj{Y_j^1=c\cdot{1+z\p_z\tilde{S}_{j,j+1}\over
1+z\p_z\tilde{S}_{0,1}}.}
This is then expressed as a power series in $q$; the results of these
computations are displayed\foot{The tables express the couplings as
series in $q^n/(1-q^n)$, from which the power series expansions themselves
are easily derived.} in tables 2 and 3 (which cover the cases
$4\le d\le10$).

\subsec{Factorization and the Other Yukawa Couplings}

Armed with the Yukawa couplings $Y^1_j$, we can give a second expression
for the $d$-point functions which were calculated in section III,
by using the factorization rules.  We first calculate
\eqn\eKNN{\eqalign{\kappa_{\alpha\alpha\cdots \alpha}&=\int\Omega\wedge\,
D_\alpha\cdots D_\alpha\Omega
=\langle {\cal O}^{(1)}\cdots {\cal O}^{(1)}\rangle  \cr
&=C^{(1,1)}\langle {\cal O}^{(2)}{\cal O}^{(1)}\cdots {\cal O}^{(1)}\rangle
=C^{(1,1)}C^{(2,1)}\langle {\cal O}^{(3)}{\cal O}^{(1)}\cdots {\cal
O}^{(1)}\rangle  \cr
&=\cdots=C^{(1,1)}\cdots C^{(d-2,1)}\langle {\cal O}^{(d-1)}{\cal
O}^{(1)}\rangle  .}}
As pointed out in equation \ealpha, we have
$C^{(i,j)}=c^{-1}Y^i_j$. Using this, and the relation
$\langle {\cal O}^{(d-1)}{\cal O}^{(1)}\rangle =c$ we find
\eqn\eKNNbis{\kappa_{\alpha\alpha\cdots \alpha}
=(c^{-1})^{d-2}\,Y^1_1Y^1_2\cdots Y^1_{d-2} .}

The $d$-point function can then be calculated from
the three-point functions given
in tables 2 and 3; when one does so, one finds precisely the same series
for $d$-point functions as given in table 1.  This remarkably delicate
factorization property of the $d$-point function power series provides
strong evidence that we have not only correctly found the coordinates to
use in the mirror map, but we have found the correct bases for the entire
horizontal subspace which maps to the integral, topological basis of
the vertical subspace under mirror symmetry.

The three-point functions $Y^1_j$ can also be used to generate other
 Yukawa couplings $Y^i_j$ with $j\ne1$, using formula \eGeneral.
We have explicitly calculated these, and displayed the answers
in tables 2 and 3 (for $4\le d\le10$) along with the $Y^1_j$'s calculated
previously.

\newsec{Mathematical Interpretation  and Comparison of Instanton Sums}

\nref\rCurve{M. Dine, N. Seiberg, X.-G. Wen, and E. Witten,
Nucl. Phys. {\bf B289} (1987) 319.}%

In the case of $d = 3$, the interpretation of the series expansion of
$\kappa_{\alpha_1 \alpha_2 \alpha_3}$
in terms of the rational curves on the mirror
is by now well known
\refs{\rCurve,\rWtop}.
For $ d \gt 3$, in addition to the existence
of more than one kind of Yukawa coupling,
there is one other important new
consideration. Holomorphic curves are no longer generically isolated
as they are for $ d = 3$, but rather come in continuous families. Thus,
the integers which arise in the series expansion to a coupling using the mirror
map no longer count numbers of curves. In this section we will give
the mathematical interpretation of the integers so found, describe
what can be calculated using more traditional mathematical methods, and
compare our results.

In the last section we gave a simple algorithm for calculating the instanton
expansions for three-point functions on certain Calabi--Yau $d$-folds.
We now seek the mathematical interpretation of the integers we have found.
The easiest way to approach this question from the physics vantage point
is to phrase our three-point functions as correlation functions in
the {\bf A} model of \rWitten.

We start with three  of our chosen basis vectors
$e_i\in H^{i,i}$, $e_j\in H^{j,j}$ and $e_{d-i-j}\in H^{d-i-j,d-i-j}$, and fix
three points $P_1=0$, $P_2=1$ and $P_3=\infty$
on the worldsheet $\Sigma=\CP1$.
We choose explicit complex submanifolds $H_i$,
$H_j$ and $H_{d-i-j}$ of complex codimension $i$, $j$, and $d-i-j$,
respectively,
which are Poincar\'e dual to the cohomology classes.  We form local
operators ${\cal O}^{(i)}(P_1)$, ${\cal O}^{(j)}(P_2)$, and
${\cal O}^{(d-i-j)}(P_3)$ which have delta function support on maps
$\Phi:\Sigma\to\tilde M$
for which $\Phi(P_1)\in H_i$ (or $\Phi(P_2)\in H_j$, or $\Phi(P_3)\in
H_{d-i-j}$
in the other cases).  The three-point function
$\langle {\cal O}^{(i)}{\cal O}^{(j)}{\cal O}^{(d-i-j)}\rangle $
can be written as a
sum over cohomology classes  of maps $\Phi$.
We index those classes by specifying $\psi$, the class of the image
of the map, and $n$, the degree of the map.  We let $\Phi_{\psi,n}$
be a typical map in its class.  Then the three-point function can
be written as \rCurve:
\eqn\eIS{
\langle {\cal O}^{(i)}{\cal O}^{(j)}{\cal O}^{(d-i-j)}\rangle
=\sum_{\psi,n} e^{\int_{\Sigma}\Phi_{\psi,n}^*(K)}\,
\#\left({\cal G}^{(i,j)}_{n,\psi}\right),
}
where
\eqn\eGee{\eqalign{{\cal G}^{(i,j)}_{n,\psi}=\{&\hbox{holomorphic maps }
\Phi:\Sigma\to\tilde{M} \hbox{ of degree $n$ and class $\psi$} \cr
&\hbox{such that }
\Phi(P_1)\in H_i, \Phi(P_2)\in H_j, \Phi(P_3)\in H_{d-i-j}\} .}}
As it stands, formula \eIS\ is somewhat problematic, since
the moduli spaces of holomorphic maps $\Sigma\to\tilde M$
which are not one-to-one (i.e., $n\ne1$) fail to have the expected dimension;
thus, the set ${\cal G}^{(i,j)}_{\psi,n}$ of maps satisfying the stated
conditions is not finite when $n\gt1$.
There is a cure for this, however, in the form of a ``multiple cover
formula'' which for threefolds was conjectured in \rCDGP\ and
proven in \rAM.  We extend this formula to the present context
in Appendix B.  Using it, we can rewrite our expression
using degree 1 maps only:
\eqn\eISone{
\langle {\cal O}^{(i)}{\cal O}^{(j)}{\cal O}^{(d-i-j)}\rangle
=\langle e_ie_je_{d-i-j}\rangle +
\sum_{\ell\gt0} {q^\ell\over 1-
q^\ell}
\, n_j^i(\ell)}
where $q=e^K$ is the single parameter, $\ell$ is the degree of a homology
class $\psi_\ell$,
 and\foot{The notation $n_j^i(\ell)$ is chosen to match that
of \rKatz.}
\eqn\enij{n_j^i(\ell)=\#\left({\cal G}^{(i,j)}_{1,\psi_\ell}\right).}
(It is necessary to separate out the degree 0 ``constant'' maps when writing
\eISone, since they are not included in the multiple cover analysis, but
lead rather to the ``topological'' term $\langle e_ie_je_{d-i-j}\rangle $.)
The entries in tables 2 and 3 have been written in the form \eISone, and it
is gratifying to observe that all calculated coefficients are in fact
integers.  These integers are predicted to coincide with the numbers
$ n_j^i(\ell)$.

\nref\rGromov{
M. Gromov,
Invent. Math. {\bf 82} (1985) 307;
Proc. Intern. Congress Math.,
Berkeley 1986, American Mathematical Society (1987) 81.}%
\nref\rTSM{E.~Witten, Commun. Math. Phys. {\bf 118} (1988)  411.}%
\nref{\rMcDuff}{D. McDuff, Invent. Math. {\bf 89} (1987) 13.}%
\nref\rSK{S. Katz, in
{\it Essays on Mirror Manifolds}, (S.-T. Yau, editor), International Press,
 1992, p. 168.}%

The actual
calculation of the numbers $n_j^i(\ell)$ using classical techniques in
algebraic geometry is a challenging
task. There are two principal difficulties.  First, the moduli spaces
${\cal G}^{(i,j)}_{1,\psi_\ell}$ may fail to have
dimension zero (even though $n=1$)
for a particular choice of complex structure on $\tilde M$. Zero-dimensional
 moduli
spaces can sometimes be obtained
by perturbing the original complex structure, but in general it is necessary to
pass to a
nearby almost-complex structure in order to guarantee the correct dimension
\refs{\rGromov,\rTSM,\rMcDuff}.
Doing so allows the number $n_j^i(\ell)$ to be calculated in principle,
but in practice it is not known how to carry out the calculation in terms of
the almost-complex structure.  Techniques for calculating $n_j^i(\ell)$
directly on $\tilde M$
(even when ${\cal G}^{(i,j)}_{1,\psi_\ell}$ has the wrong dimension)
have been pioneered by Katz \rSK, but these techniques do not yet apply
in complete generality.

The second difficulty occurs even when no perturbation of complex structure
is necessary.  Simply put, the evaluation of the numbers $n_j^i(\ell)$
using the
classical tools of algebraic geometry is a very hard task, and effective
methods are not known except in the simplest cases.  To calculate
$n_j^i(\ell)$,
one first describes ${\cal G}^{(i,j)}_{1,\psi_\ell}$ as an intersection
of certain subvarieties in a moduli space of curves.  (This is the translation
of \eGee\ into algebraic geometry.)  The number of points in the space
should then be found using the standard techniques of
algebraic intersection theory.  However, those techniques require a compact
moduli space, and the moduli space at hand is not compact.  It can be
compactified by adjoining points corresponding to certain ``limiting''
curves of other types---the resulting compact space is known as a Hilbert
scheme.  The delicate part of the computation is to properly account for
the portion of the answer which comes from the limiting curves, and
this requires knowing the structure of those curves in detail.  As $\ell$
increases, the types of limiting curves which must be considered
grow more and more complex.

For $\ell=1$ and $2$, these difficulties can be overcome, and Katz \rKatz\
has checked the predictions in tables $2$ and $3$ for $\ell=1$ and $2$
(that is, the coefficients of $q\over1-q$ and $q^2\over1-q^2$),
obtaining agreement in each case.\foot{Very recently Ellingsrud and Str{\o}mme
have also verified some of our predictions for $\ell=3$  \rESprivate.}

The associativity relations \eGeneral\ now imply some relations among the
numbers $n_j^i(\ell)$ which had not been observed in the mathematics
literature.
It is likely that the geometric explanation of these relations in terms
of four-point functions which has been put forward by Witten \rW\ can
be used to give a complete mathematical proof of these new relations.
(The subtleties in that proof would again involve issues of compactifying
moduli spaces appropriately.)  Katz \rKatz\ has directly proved these
relations in the case $\ell=1$.

\newsec{Conclusions}

Our focus in this paper has been an analysis of some aspects of
mirror symmetry for Calabi--Yau manifolds whose complex dimension
is greater than three, the previously studied case.
We have found that a number of new issues arise. First, the geometric
constraints characterizing the associated complex structure and K\"ahler
moduli spaces differ from the threefold case, in which they
have usually been referred
to as the ``constraints of special geometry''. The analogue of
special geometry in
the higher dimensional case (for one-parameter families)
can be summarized by a general constraint
valid for all dimensions including three---equation \eRiem---but
the explicit evaluation of this constraint in terms of the Riemann
curvature tensor and the Yukawa couplings is certainly sensitive
to the dimension. We have explicitly worked this out for one-parameter
examples in the case of dimension four and five. Second, whereas
there is one type of Yukawa coupling (in each of the $\bf A$ and
$\bf B$ models) in the case of dimension three, the number of Yukawa
couplings rapidly grows as a function of the dimension. By making use
of the associativity of the operator product algebra, we
identified a fundamental subset of couplings on which all others are
functionally dependent. Third, whereas the exploitation of mirror symmetry
in the case of threefolds only requires understanding a preferred set of
moduli space coordinates (``special coordinates''), in higher dimension
we require more structure: a preferred basis of (part of) the cohomology
ring. We have presented an efficient algorithm for generating such
bases (in one-parameter models), making use of
the Gauss--Manin connection. Furthermore, we have
shown that our procedure naturally reproduces the special coordinates
discussed in the three dimensional setting as well as giving a calculationally
tractable procedure for generating the independent set of Yukawa couplings.
Fourth, in dimension three, rational curves on a Calabi--Yau manifold
are generically isolated whereas in higher dimension they come in families.
This requires a reinterpretation of the instanton expansion of
Yukawa couplings in higher dimension in terms of the characteristic
classes of the parameter spaces of rational curves. We  have done this
and explicitly carried out such calculations for one-parameter
Calabi--Yau manifolds of complex dimension at most ten.
In the limited number of cases in which such characteristic
classes can be effectively calculated using conventional mathematical
methods, we find agreement.
The calculational power of mirror symmetry is thereby once again
affirmed.

\bigbreak\bigskip\bigskip\centerline{{\bf Acknowledgements}}\nobreak
We thank
Sheldon Katz for useful discussions,  particularly for the zest
with which he tackled the task of verifying (and helping us to
refine) our predictions.  We also thank Cumrun Vafa for communicating
the results of \rBCOV\ to us prior to publication.
The work of B.R.G.\ was supported
by a National Young Investigator award, by the Ambrose Monell Foundation and
by the Alfred P. Sloan Foundation,
the work of D.R.M.\ was supported by NSF grants DMS-9103827 and DMS-9304580
and by an American Mathematical Society Centennial Fellowship,
and the work of M.R.P.\ was supported  by DOE grant DE-AC02-76ER03075,
by a Presidential Young Investigator Award, by NSF grant
PHY 92-45317 and by the W. M. Keck Foundation.

\appendix{A}{Some Remarks on Covariant Derivatives}

The analysis of section II involved several times a need to
differentiate sections of (non-holomorphic)
bundles of the form $V =
 {\cal H} \otimes {\cal L }^{-1} \otimes (T^*)^p \otimes
(\overline T^*)^q \otimes {\overline {\cal H }}\otimes
\overline {\cal L}{}^{-1}$,
where $T$ is the
holomorphic tangent bundle of the moduli space $\cal M$.
Since each factor occurring in this bundle is itself either holomorphic or
antiholomorphic there is a natural connection we can define on the
tensor product.
Namely, on each component factor we define the complex metric connection
and we extend this connection to the product by the Leibnitz rule.
More specifically, if $Q$ is a holomorphic bundle with Hermitian fiber
metric $h_{a \overline b}$, there is a unique connection which is
compatible with the metric, i.e.
\eqn\eCom{ d \langle \overline s  \vbar  t\rangle  = \langle \overline {Ds}
\vbar  t\rangle  + \langle \overline
s  \vbar  Dt\rangle }
where $s$ and $t$ are local smooth sections of $Q$ and the inner product
$\langle   \vbar  \rangle $ is that given by $h$, and which agrees with
ordinary $\overline \del$
differentiation in the $(0,1)$ direction.  The connection $\omega$
satisfying these conditions can be written
\eqn\eConn{ \omega = (\del h) h^{-1}.}
Clearly this construction also works for an antiholomorphic bundle by
demanding agreement with partial differentiation in the $(1,0)$ direction.
(One must take the complex conjugate of the formulas.)
Quite generally, if we have connections on each of $n$ bundles $A_1,\ldots ,
A_n$, then the sum of these connections provides a connection on the
product bundle $A_1 \otimes\cdots\otimes A_n$. Hence, by using the complex
metric connections
or their complex conjugates
on each individual factor, their sum is a connection
on $V$. Of course, this connection, while compatible with
the metric, no longer agrees with partial differentiation in either the
$(1,0)$ or $(0,1)$ directions.

It proves instructive
to explicitly write out one consequence of metric compatibility. Let
$s$ and $t$ be sections of $V$. Metric compatibility implies
\eqn\eComp{
\eqalign{
 d \langle \overline s  \vbar  t\rangle  & = \del_{\alpha} \langle \overline s
\vbar  t\rangle
dz^{\alpha} +
                                  \overline \del_{\overline \alpha} \langle
\overline
                                  s  \vbar  t\rangle  d \overline z^{\overline
\alpha}\cr
                     &  =  \langle \overline {Ds}  \vbar  t\rangle  + \langle
\overline s  \vbar
Dt\rangle \cr
                     &  =  \langle \overline{(D^{1,0} + D^{0,1})s}  \vbar
t\rangle
                            + \langle \overline s  \vbar  (D^{1,0} + D^{0,1})
t\rangle . \cr
}}
Now, we can
decompose the equality above by type to get
\eqn\eIBP{ \del_{\alpha} \langle  \overline s  \vbar  t\rangle  =
\langle \overline{D^{0,1}_{\overline
\alpha} s}  \vbar  t\rangle  + \langle \overline s  \vbar  D^{1,0}_{\alpha}
t\rangle }
and its complex conjugate. From \eIBP\ we then have
\eqn\eIBPP{\del_{\alpha} \langle  \overline s  \vbar  t\rangle  =
\langle D^{1,0}_{\alpha} \overline s  \vbar  t\rangle
+ \langle \overline s  \vbar D^{1,0}_{\alpha} t\rangle }
where $D^{1,0}$ in the first term on the right hand side is a covariant
derivative acting on sections of $\overline V$.

Implicit in the above discussion is that the symbol $\langle   \vbar  \rangle $
is the inner
product on $V$. More generally, we can replace this inner product on
$V$ by an inner product just on ${\cal H } \otimes {\cal L}$,
 $\langle   \vbar  \rangle _{ {\cal H } \otimes {\cal L}}$.
Then, $\langle \overline s  \vbar  t\rangle _{{\cal H }\otimes {\cal L}}$ is a
section of
$(T^*)^p \otimes \overline {(T^*)^q}$
 and we similarly have
\eqn\eIBPF{D^{1,0}_{\alpha} \langle \overline s  \vbar  t\rangle _{ {\cal H }
\otimes {\cal L}} =
\langle D^{1,0}_{\alpha} \overline s  \vbar  t\rangle  + \langle \overline s
\vbar  D^{1,0}_{\alpha}
t\rangle .  }
It is important to bear in mind that in \eIBPF\ the meaning of the derivative
is determined by the object on which it acts. Explicitly, the $D^{1,0}$
on the left hand side acts on sections of $(T^*)^p \otimes
\overline {(T^*)^q}$; the first on the right hand side acts on sections
of $\overline V$ while the last acts on sections of $V$.
We have repeatedly made use of \eIBPF\ in section II.

\appendix{B}{The Multiple Cover Formula in Higher Dimension}

Let $X$ be a Calabi--Yau $d$-fold.
Our derivation of the multiple cover formula roughly follows section 4 of
\rAM, but there are some new twists in higher dimension.
We let ${\cal H}$ be a fixed
component of the Hilbert scheme of $X$, which parametrizes a family
$f:{\cal C}\to{\cal H}$ of rational curves on $X$ with the property
that $T_X|_C=
{\cal O}_C(2)\oplus{\cal O}_C(-1)\oplus
{\cal O}_C(-1)\oplus{\cal O}_C^{\oplus(d-3)}$
for the general curve $C$ in the family.
Let $M_n^{\cal H}$ be the moduli space for holomorphic maps $\Phi:\CP1\to X$
which are  degree $n$ covers of the rational curves parametrized by
${\cal H}$.
We wish to evaluate the contributions to three-point functions
$\langle {\cal O}^{(i)}{\cal O}^{(j)}{\cal O}^{(d-i-j)}\rangle $
made by $M_n^{\cal H}$.  Since $M_n^{\cal H}$ has the wrong dimension,
we will need to calculate the top Chern class of a certain bundle.

There is a natural
embedding $\iota:{\cal C}\to X$.
We can describe the family ${\cal C}$ in terms of the sheaf
${\cal V}:=f_*{\cal O}_{\cal C}(1)$.  Over a Zariski-open subset
${\cal H}_0\subset{\cal H}$, this sheaf restricts to a locally free
sheaf ${\cal V}_0$ of rank two, and the
 $\CP1$-bundle
$\IP({\cal V}_0)$ is birational to ${\cal C}$.

By blowing up ${\cal H}$, we may assume that ${\cal V}_0$ has a locally
free extension.  We shall do this, and shall also replace ${\cal C}$
by the projectivization of that locally free extension.  After making
those birational modifications of our data, we arrive at the situation
in which ${\cal V}=f_*{\cal O}_{\cal C}(1)$ is locally free, and
${\cal C}=\IP({\cal V})$.  The modifications we have made can be expected
to be located outside of the subspace in which the calculation of the
 three-point
functions is localized.  We treat the pullback $\iota^*(T_X)$ of $T_X$
to ${\cal C}$
as coinciding with ${\cal O}_{\cal C}(2)\oplus{\cal O}_{\cal C}(-1)\oplus
{\cal O}_{\cal C}(-1)\oplus{\cal O}_{\cal C}^{\oplus(d-3)}$.  This also
holds generically, and the places where it fails can be expected to be
located outside of the crucial subspace.

To describe a point in $M_n^{\cal H}$, we must specify the image curve,
and specify
a ratio of two relatively prime degree $n$ polynomials to define the map.
We compactify the moduli space using graphs of maps,
motivated by
the work of Gromov \rGromov\
(cf.\ also \rAM).
To construct the graph compactification, we first extend from pairs of
relatively prime polynomials to arbitrary pairs of polynomials, obtaining
the space
$\overline{M}:=\IP(\Sym^n{\cal V}\oplus\Sym^n{\cal V})$.
The graphs of the maps can then be naturally taken in the space
\eqn\eGraph{
Z:=\CP1\times({\cal C}\times_{\cal H}\overline{M})
}
with the closure $\overline{\Gamma}$ of the universal graph $\Gamma$
described by the equation
\eqn\eUniversal{
{s\over t}={\sum a_ix^iy^{n-i}\over\sum b_ix^iy^{n-i}},
}
or equivalently
\eqn\eUni{
t\,\sum a_ix^iy^{n-i}-s\,\sum b_ix^iy^{n-i}=0,
}
where $[x,y]$ are homogeneous coordinates on $\CP1$,
$[s,t]$ are homogeneous coordinates on a fiber $C$ of ${\cal C}$,
and $[a_0,\dots,a_n,b_0,\dots,b_n]$ give the coordinates in a fiber
of $\overline{M}\to{\cal H}$.  Counting degrees in \eUni, it
 follows that the line bundle
associated to $\overline{\Gamma}$ can be written as
\eqn\eLinebundle{
{\cal O}(\overline{\Gamma})=
\mu^*({\cal O}_{\CP1}(n))\otimes
\nu^*({\cal O}_{\cal C}(1))\otimes
\pi^*({\cal O}_{\overline{M}}(1)),
}
where $\mu:Z\to\CP1$, $\nu:Z\to{\cal C}$ and $\pi:Z\to\overline{M}$ are
the natural projection maps.

\nref\rNmatrix{E. Witten, Nucl. Phys. {\bf B371} (1992) 191.}%

The tangent bundle $T_X$ determines a bundle
${\cal E}:=(\iota\circ\nu)^*(T_X)$ on $Z$, which restricts to the
bundle ${\cal E}|_{\overline{\Gamma}}$ on the graph-closure
$\overline{\Gamma}$.
Following the methods developed in \rNmatrix\ and \rAM, we must calculate
the top Chern class of the bundle $R^1\pi_*({\cal E}|_{\overline{\Gamma}})$
whose fibers are the obstruction groups for the moduli problem.  We will
do this by using the short exact sequence
\eqn\eSES{
0\longrightarrow{\cal E}(-\overline{\Gamma})\longrightarrow{\cal E}
\longrightarrow{\cal E}|_{\overline{\Gamma}}\longrightarrow0 .
}

It is convenient to write
${\cal E}(-\overline{\Gamma})={\cal F}\otimes
\pi^*({\cal O}_{\overline{M}}(-1))$.
Then we have
\eqn\eBundles{\eqalign{&
{\cal E}=\nu^*\left(
{\cal O}_{\cal C}(2)\oplus{\cal O}_{\cal C}(-1)^{\oplus2}
\oplus{\cal O}_{\cal C}^{\oplus(d-3)}\right)
\cr
&
{\cal F}=\mu^*\left({\cal O}_{\CP1}(-n)\right)\otimes\nu^*\left(
{\cal O}_{\cal C}(1)\oplus{\cal O}_{\cal C}(-2)^{\oplus2}
\oplus{\cal O}_{\cal C}(-1)^{\oplus(d-3)}\right).
\cr}}
We compute the cohomology of these bundles on a fiber $S$ of $\pi$.
Such a fiber can be written in the form $S=\CP1\times C$, with $C$
the image of the corresponding map (one of the curves in the family
${\cal C}$).
When restricted to $S$, our bundles become
\eqn\eRestricted{\eqalign{&
{\cal E}|_S=(\nu|_S)^*\left(
{\cal O}_{ C}(2)\oplus{\cal O}_{ C}(-1)^{\oplus2}
\oplus{\cal O}_{ C}^{\oplus(d-3)}\right)
\cr
&
{\cal F}|_S=(\mu|_S)^*\left({\cal O}_{\CP1}(-n)\right)\otimes(\nu|_S)^*\left(
{\cal O}_{ C}(1)\oplus{\cal O}_{ C}(-2)^{\oplus2}
\oplus{\cal O}_{ C}(-1)^{\oplus(d-3)}\right).
\cr}}
It is easy to calculate the spaces of global sections:
\eqn\eHzero{\eqalign{&
H^0(S,{\cal E}|_S)=H^0(C,
{\cal O}(2)\oplus{\cal O}(-1)^{\oplus2}
\oplus{\cal O}^{\oplus(d-3)})\cong\IC^d
\cr
&
H^0(S,{\cal F}|_S)=\{0\}.
\cr}}
We can also compute $H^2$'s using Serre duality and the canonical bundle
formula
\eqn\eKS{K_S=(\mu|_S)^*\left({\cal O}_{\CP1}(-2)\right)\otimes(\nu|_S)^*\left(
{\cal O}_{ C}(-2)\right).}
The results are that $H^2(S,{\cal E}|_S)^*$ is isomorphic to
\eqn\eHtwoE{\eqalign{
& H^0(S,(\mu|_S)^*({\cal O}_{\CP1}(-2))\otimes
(\nu|_S)^*(
{\cal O}_{ C}(-4)\oplus{\cal O}_{ C}(-1)^{\oplus2}
\oplus{\cal O}_{ C}(-2)^{\oplus(d-3)})) \cr
&\quad = \{0\}
}}
and  that $H^2(S,{\cal F}|_S)^*$ is isomorphic to
\eqn\eHtwoF{\eqalign{
& H^0(S,
(\mu|_S)^*({\cal O}_{\CP1}(n-2))\otimes(\nu|_S)^*(
{\cal O}_{ C}(-3)\oplus{\cal O}_{ C}^{\oplus2}
\oplus{\cal O}_{ C}(-1)^{\oplus(d-3)})
\cr &\quad\cong H^0(S,(\mu|_S)^*({\cal O}_{\CP1}(n-2)\oplus
{\cal O}_{\CP1}(n-2)))\cong\IC^{2n-2} .
\cr}}

This last calculation can be done as a bundle calculation, not just
fiber by fiber.  Doing so gives a natural isomorphism
between $(R^2\pi_*{\cal F})^*$, and $R^0\pi_*{\cal G}$, where
\eqn\ecanG{
{\cal G}:=\mu^*({\cal O}_{\CP1}(n-2)\oplus
{\cal O}_{\CP1}(n-2)) .}
Because $Z$ is a product
of $\CP1$ and ${\cal C}\times_{\cal H}\overline{M}$, the bundle
$R^0\pi_*{\cal G}$
is trivial, being canonically isomorphic to
\eqn\ecanon{
{\cal O}_{\overline{M}}\otimes\,H^0(\CP1,{\cal O}_{\CP1}(n-2)\oplus
{\cal O}_{\CP1}(n-2)).}

To complete our calculation, we note that
 Riemann--Roch tells us that $\chi(S,{\cal E}|_S)=d$ and therefore
that $h^1(S,{\cal E}|_S)=0$.
As a consequence, we find that $R^1\pi_*{\cal E}=R^2\pi_*{\cal E}=0$,
and that $R^0\pi_*{\cal E}$ is locally free of rank $d$.  We also find
that the short exact sequence
\eSES\
gives rise to a long exact sequence whose nonzero terms split into two
exact sequences:
\eqn\eLES{\eqalign{
0\longrightarrow R^0\pi_*{\cal E}
\longrightarrow R^0\pi_*({\cal E}|_{\overline{\Gamma}})
& \longrightarrow R^1\pi_*({\cal E}(-\overline{\Gamma}))\longrightarrow
0, \cr
0
\longrightarrow R^1\pi_*({\cal E}|_{\overline{\Gamma}})
\longrightarrow & R^2\pi_*({\cal E}(-\overline{\Gamma}))\longrightarrow
0.
}}
It then follows from the projection formula that
\eqn\eFind{\eqalign{
R^1\pi_*({\cal E}|_{\overline{\Gamma}})&\cong
R^2\pi_*({\cal E}(-\overline{\Gamma}))\cong
(R^2\pi_*{\cal F})\otimes{\cal O}_{\overline{M}}(-1)\cong
(R^0\pi_*{\cal G})^*\otimes{\cal O}_{\overline{M}}(-1)\cr&\cong
H^0(\CP1,{\cal O}_{\CP1}(n-2)\oplus
{\cal O}_{\CP1}(n-2))^*\otimes{\cal O}_{\overline{M}}(-1) .
}}

We now see that the top Chern class
$c_{2n-2}(R^1\pi_*({\cal E}|_{\overline{\Gamma}}))$
coincides with
\eqn\etop{
c_1({\cal O}_{\overline{M}}(-1))^{2n-2}
=c_1({\cal O}_{\overline{M}}(1))^{2n-2} ,}
 and so is a class whose
intersection with every fiber of $\overline{M}\to{\cal H}$ is
a linear space of dimension $3$.

The contribution of $M_n^{\cal H}$ to the three-point function
$\langle {\cal O}^{(i)}{\cal O}^{(j)}{\cal O}^{(d-i-j)}\rangle $
is calculated by an integral
\eqn\eintegral{
\int_{\overline{M}} \overline{e}_i \wedge \overline{e}_j
\wedge \overline{e}_{d-i-j} \wedge
c_{2n-2}(R^1\pi_*({\cal E}|_{\overline{\Gamma}})) ,}
where the $\overline{e}$'s are the induced classes on $\overline{M}$,
with delta-function support on those maps which take a fixed basepoint $P$ to
a fixed cycle $H$.  These integrals localize on a finite number of fibers
of $\overline{M}\to{\cal H}$, and in each such fiber the last term
$c_{2n-2}(R^1\pi_*({\cal E}|_{\overline{\Gamma}}))$ serves to reduce
the integral to an integral over $\CP3$.
Each delta-function support condition has the same cohomological effect
on $\CP3$ regardless of the value of $n$, so we recover the {\it same}\/
instanton contribution for $n\gt1$ as for $n=1$, namely, the
number of points in ${\cal H}$ whose corresponding rational curve
meets the stated conditions.  Summing over $n$,
we get a term of the form $q^{\ell}/(1-q^{\ell})$ times the $n=1$
instanton number, as asserted in \eISone.

\listrefs

\bye